\documentclass[reprint,
superscriptaddress,
nofootinbib,
 amsmath,amssymb,
 aps,pra
]{revtex4-2}
\usepackage{graphicx}% Include figure files
\usepackage{dcolumn}% Align table columns on decimal point
\usepackage{bm}% bold math
\usepackage{xcolor}
\usepackage{ulem}
\usepackage{chemformula} % Formula subscripts using \ch{}

\usepackage{textgreek}

\usepackage[separate-uncertainty=true]{siunitx}
\DeclareSIUnit\Molar{\textsc{M}}
\begin{document}

\preprint{APS/123-QED}

% \title{An optomechanical coin flip, wavelength controlled rotations in an erbium-doped levitated system}% 

\title{An Optomechanical Coin Flip: Wavelength-Modulated, Erbium-Powered Rotations in a Levitated System}

%"Wavelength controlled optically levitated gyroscopes exhibiting tennis racket effects" OR "Optically levitated gyroscopes with wavelength driven rotational control facilitated by an atomic erbium dopant" OR
% OR "A red-blue spinning top - modifying the free body rotations of levitated micron scale discs by pumping electronics states of Erbium

% particles using optical radiation}
%\title{Force/position detection limit with an optically levitated high aspect ratio disc in the Mie-Lorentz regime.}% Force line breaks with \\

% \author{George Winstone}
% \altaffiliation{These authors contributed equally}
% \affiliation{Center for Fundamental Physics, Department of Physics and Astronomy,Northwestern University, Evanston, Illinois 60208, USA}

% \author{First author (this should be an above board democratic decision between authors) - Please re order yourselves after this as you see fit}
\author{George Winstone}
\affiliation{Center for Fundamental Physics, Department of Physics and Astronomy,Northwestern University, Evanston, Illinois 60208, USA}
\affiliation{Center for Interdisciplinary Exploration and Research in Astrophysics, Department of Physics and Astronomy, Northwestern University, Evanston, Illinois 60208, USA}

\author{Maddox Wroblewski}
\affiliation{Center for Fundamental Physics, Department of Physics and Astronomy,Northwestern University, Evanston, Illinois 60208, USA}
\affiliation{Department of Physics, The University of Texas at Austin, Austin, TX 78712, USA}

\author{Lars Forberger}
\affiliation{Department of Materials Science \& Engineering, University of Washington, Seattle, Washington 98195, USA}

\author{Zhiyuan Wang}
\affiliation{Center for Fundamental Physics, Department of Physics and Astronomy,Northwestern University, Evanston, Illinois 60208, USA}

\author{Shelby Klomp}
\affiliation{Center for Fundamental Physics, Department of Physics and Astronomy,Northwestern University, Evanston, Illinois 60208, USA}

\author{Scott Grudichak}
\affiliation{Center for Fundamental Physics, Department of Physics and Astronomy,Northwestern University, Evanston, Illinois 60208, USA}

\author{Shafaq Gulzar Elahi}
\affiliation{Center for Fundamental Physics, Department of Physics and Astronomy,Northwestern University, Evanston, Illinois 60208, USA}

\author{Yuqi Qian}
\affiliation{Center for Fundamental Physics, Department of Physics and Astronomy,Northwestern University, Evanston, Illinois 60208, USA}
\affiliation{Center for Interdisciplinary Exploration and Research in Astrophysics, Department of Physics and Astronomy, Northwestern University, Evanston, Illinois 60208, USA}

\author{Miriam M Flórez}
\affiliation{IQQI, Innsbruck, Austria, EU}

% \author{Tracy Northrup}
% \affiliation{IQQI, Innsbruck, Austria, EU}

\author{Zhaojie Feng}
\affiliation{Department of Materials Science \& Engineering, University of Washington, Seattle, Washington 98195, USA}

\author{Peter J. Pauzauskie}
\affiliation{Department of Materials Science \& Engineering, University of Washington, Seattle, Washington 98195, USA}
\affiliation{Physical and Computational Sciences Directorate, Pacific Northwest National Laboratory, Richland, Washington 99352, USA}

\author{Andrew A. Geraci}
\affiliation{Center for Fundamental Physics, Department of Physics and Astronomy,Northwestern University, Evanston, Illinois 60208, USA}
\affiliation{Center for Interdisciplinary Exploration and Research in Astrophysics, Department of Physics and Astronomy, Northwestern University, Evanston, Illinois 60208, USA}

%\collaboration{and the LSD Collaboration}%\noaffiliation

\date{\today}% It is always \today, today,

\begin{abstract}
Optical levitation of nano-scale systems offers a pathway to highly sensitive rotation measurements, which are critical for advancing gyroscopic technologies. While prior studies have primarily focused on controlling rotational degrees of freedom of optically levitated particles via modulation of optical power, polarization, and ambient pressure, here we demonstrate wavelength-controlled rotation of the ``coin-flip'' mode in optically levitated $\text{NaYF}$ hexagonal prisms doped with erbium by modulating the wavelength of a secondary pump beam. By switching the pump light wavelength, we precisely modulate the particle’s rotation rate in a binary fashion, encoding the ASCII message “hello” in its rotational frequency. Finally, we observe long-term bimodal periodic dynamics in the rotational motion of a levitated prism that are suggestive of a Dzhanibekov (or tennis-racket)-like effect.

%%% Alternative abstract:

%“We demonstrate the first wavelength-based control of rotation in an optically levitated system, using Er-doped NaYF₄ hexagonal prisms. By modulating the pump wavelength between 1535 nm and 1565 nm, the rotation rate switches between two discrete values, enabling binary encoding of a 40-bit ASCII message (‘HELLO’) in the mechanical motion. We also observe long-term bimodal flipping dynamics consistent with a classical Dzhanibekov (‘tennis-racket’) instability. This establishes a new optical control handle for gyroscopic and quantum-rotational applications.”

\end{abstract}

\maketitle

\section{Introduction}

Rotational effects in levitated optomechanics have grown into a topic of significant interest to the scientific community over recent years \cite{zeng2024optically} for a variety of reasons including gyroscopic navigation, biological analysis \cite{rademacher2025roto}, observing the Casimir torque \cite{xu2017detecting,Ju2023}, pressure sensing, and the possibility to extend the boundaries of fundamental physics by observing decoherence-resilient quantum dynamics of a macroscopic system \cite{stickler2018probing}. Indeed, the fastest-spinning man-made objects are optically levitated nanoparticles \cite{reimann2018ghz,Li2018}.

%free rotations

Additionally, significant scientific effort has been expended in recent years on the production of massive %(in terms of mass(kg)) 
 quantum superposition states. Wave-like behavior and matter-wave interference has been experimentally demonstrated for electrons ~\cite{Tonomura:1989}, neutrons ~\cite{Colella:1975dq,Werner:1979gi,Rauch:2015jkh}, atoms ~\cite{Fixler:2007is,Asenbaum:2016djh,Overstreet:2021hea,Amit,SGI_experiment}, and even complex molecules and clusters \cite{Eibenberger:2013cqb,Fein:2019dgf,pedalino2025probingquantummechanicsusing}, and levitated optomechanical systems represent a promising platform to achieve delocalized quantum states and wave-like behavior in systems with several orders of magnitude larger mass \cite{ORI11_GM,Schut:2024lgp,Bateman2014,Goldman2015}. Levitated nanoparticles have been able to be cooled to the quantum ground state of their trapping potential for their translational degrees of freedom ~\cite{delic2020cooling,tebbenjohanns2019damping,kamba2022optical,ranfagni2022twodimensional}, librational degrees of freedom \cite{Dania2025}, and recent work has shown a quantum wave packet delocalization beyond the size scale of a nanoparticle \cite{Rossi2025}. Decoherence effects represent one of the greatest challenges in achieving macroscopic spatial superpositions with large masses \cite{nimmrichter2014macroscopic}. A number of methods to deal with decoherence effects in massive macroscopic state generation have been proposed. These include performing the interference quickly  %such as the SUPER-MARIO protocol
 \cite{neumeier2024fast}, or performing the interference at a very low temperature to suppress decoherence effects \cite{nimmrichter2014macroscopic}.

In this context, degrees of freedom that are resistant to decoherence effects are especially valuable. In particular, quantum mechanical equivalents of free rotations are expected to display strong resilience to classical noise channels \cite{stickler2018probing}. These advantages have also been projected to be observed for the quantum tennis-racket-effect \cite{ma2020quantum}. Notably, the appearance of quantum dynamics in the tennis-racket-effect is independent of the rotor mass \cite{ma2020quantum}, potentially allowing for the creation of larger quantum superpositions than when created via linear translational states. Recent experimental work in cooling the librational degrees of freedom of an optically levitated particle \cite{gao2024feedback} specifically %mentions 
targets the quantum tennis-racket-effect as a goal, making this exploration of such a topic in an optically levitated system a timely endeavor for the community. %addition to the literature. 

In the macroscopic world, torsion balances are some of the most sensitive experiments constructed on a tabletop and provide some of the best limits in setting bounds on existing physical laws and exploring new novel physics
%\cite{westphal2021measurement}\cite{fleischer2022cryogenic}
\cite{westphal2021measurement, fleischer2022cryogenic}. In the world of nano- and micro-scale levitated optomechanics, the same fundamental principle is possible at orders of magnitude lower length scale, with the librational bounds of an optically levitated microparticle providing the arms of the torsion balance \cite{ahn2020ultrasensitive}.

% We love Gyroscopes (decoherence tolerant quantum revivals for matterwave interferometery\cite{stickler2018probing}, exploring casimir torque, applications in navigation etc).

Gyroscopes, particularly optically levitated nanoparticle-based gyroscopes, have garnered significant attention in recent research \cite{zeng2024optically} with the potential for parts-per-billion (ppb) scale factor stability gyroscopes via optical levitation of nanospheres and microspheres. One of the requirements is a high moment of inertia, low-noise levitator with center-of-mass cooling that can be rotated extremely quickly in free space while decoupled from its environment \cite{zeng2024optically}. Optically levitated nanoparticles also show very low damping rates when spun around the most symmetric axis \cite{zielinska2024long}. Together, these factors of high rotational speeds, low damping, and the potential to feedback cool the non-rotating degrees of freedom to ground state (thus reducing the rotor noise from the gyroscope perspective) suggest a potentially rich platform for gyroscope development. Our experimental work in this paper %directly 
adds another %handle 
method into the toolbox for controlling the rotational states of optically levitated systems.

Finally, to place our work into additional context, significant effort has recently gone into engineering `smart particles,' which are built or grown in bespoke topologies and shapes, and in some cases have additional interesting and useful internal degrees of freedom added via doping. One such example is \ch{\textbeta-NaYF} hexagonal prisms which have been optically levitated \cite{winstone2022optical, aggarwal2022searching} and can be doped with lanthanides such as erbium or ytterbium to allow for solid state cooling
%\cite{rahman2017laser}\cite{laplane2024inert}
\cite{rahman2017laser, laplane2024inert}. Similar flat geometries, which may prove to be advantageous for overcoming photon recoil heating \cite{aggarwal2022searching}, might also be amenable to printing top-down degrees of freedom with classic photolithography techniques, potentially allowing for a convergence of top-down and bottom-up added degrees of freedom in a levitated system. 

In this work, we introduce control of the rotational frequency of a \SI{5}{\percent} Er:\ch{\textbeta-NaYF} levitated system by tuning the wavelength of a pump laser. This is enabled through use of the erbium dopant as an atomic engine to facilitate the driving/control. Rotational frequency control like this has only previously been achieved through tuning of optical power and/or polarization (utilizing the objects shape \cite{Kuhn:17}, or birefringence \cite{arita2023cooling}, or structured light with orbital angular momentum \cite{hu2023structured}). 

% This additional method of control gives system designers more `handles' on rotation speed for applications in constructing PID loops out of rotational optomechanical systems (existing studies have focused mainly on power and polarisation based control). This allows for potential simultaneous decoupled control over translational frequencies (via power) and rotational frequencies (via pump wavelength) at the same time.

Using wavelength as an additional control parameter offers potential advantages over intensity and/or polarization tuning. Wavelength modulation can in principle adjust the rotational frequency without altering the total optical power at the trap, thereby avoiding unwanted changes to translational confinement. Wavelength based control also does not have the requirement of using a birefringent material or appropriate geometry amenable to polarization based control.

% \fixme{[AG: to better motivate the significance of the results, can we elaborate or say something further about possible advantages of using wavelength to tune rotation rather than intensity or polarization?]}

% \textcolor{blue}{"Using wavelength as a additional control parameter offers several advantages over intensity and/or polarization tuning. First, wavelength modulation can adjust the rotational frequency without altering the total optical power at the trap, thereby avoiding unwanted changes to translational confinement. }

% \textcolor{purple}{GW: How about the following::: "Using wavelength as a additional control parameter offers several advantages over intensity and/or polarization tuning. First, wavelength modulation can adjust the rotational frequency without altering the total optical power at the trap, thereby avoiding unwanted changes to translational confinement. Second, unlike polarization-based control—which is often limited by birefringence, or particle geometry—the wavelength. Finally, this scheme introduces a path to orthogonal control channels in which translational and rotational degrees of freedom can be independently stabilized, enabling higher-bandwidth feedback and ultimately providing significant potential for improving scale-factor stability in optical levitation based rotation sensors and gyroscopes."}

\begin{figure}
    \centering
    \includegraphics[width=0.44\textwidth]{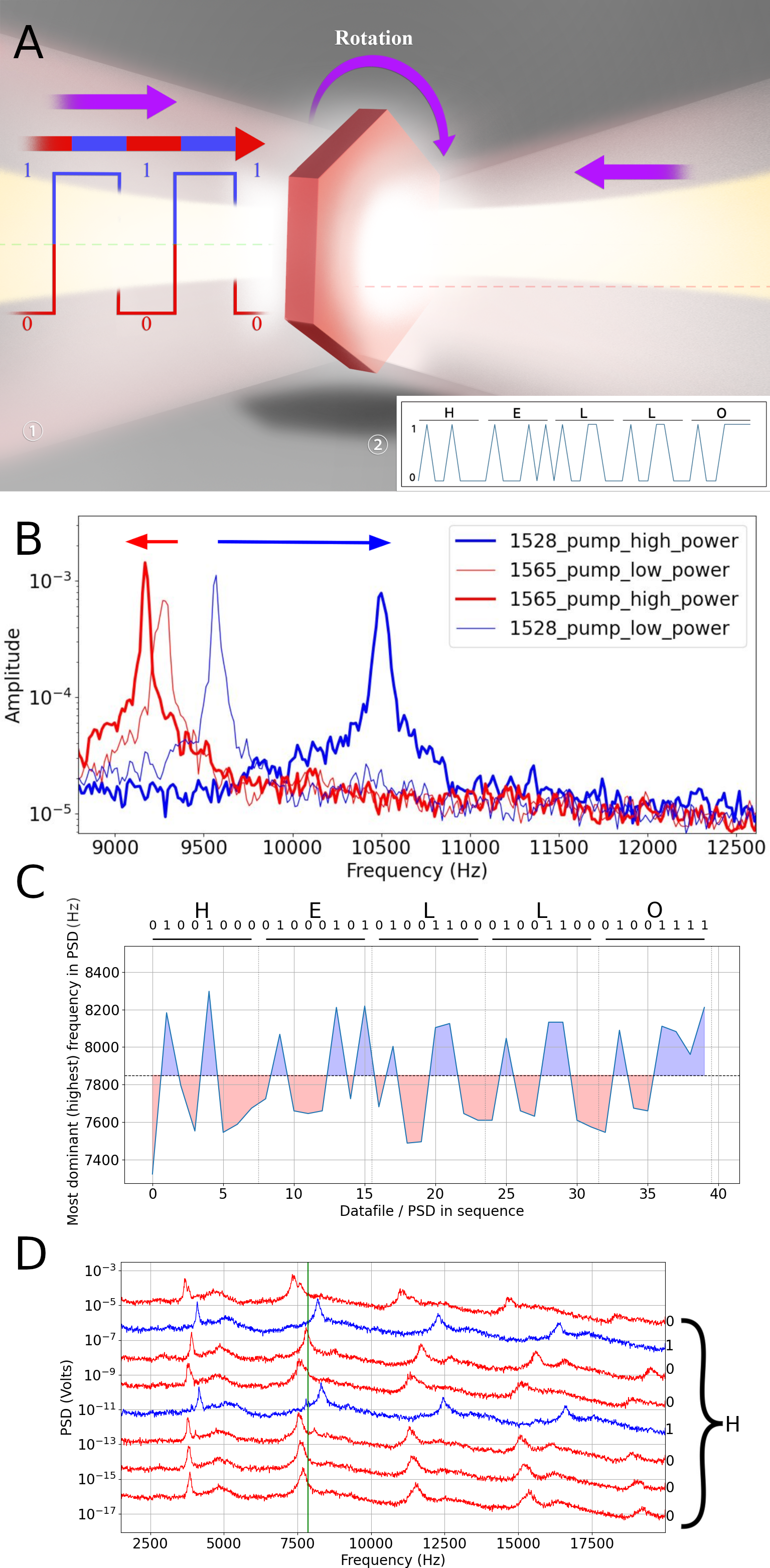}
    \caption{\textbf{A}: Conceptual figure of an optically levitated hexagon receiving a bitwise binary control signal in the form of red- and blue-detuned (with respect to the trapping light) pumping laser, which causes the hexagon to rotate faster or slower depending on the wavelength of the control signal. Blue causes the hexagon to rotate faster while red either causes less change or causes it to slow down depending on the background pressure of the system. \textbf{B}: The frequency domain data of the free rotational motion of an optically levitated hexagon when being pumped with red-detuned and blue-detuned light at \SI{2}{mbar} pressure. This data is zoomed in to the region of interest of the highest amplitude peak in the Fourier spectrum of the optically levitated test mass's motion. Here low power indicts \SI{0.1}{mW} of pump power and high indicates \SI{31}{mW} of pump power. \textbf{C}: Hello world sequence generated by changing the wavelength of the applied pump beam (blue pump = 1, red pump = 0). Encoded ASCII sequence is sufficiently resilient to long term drifts to encode a 40-bit message in wavelength space, which is then read out in the mechanical motion of the particle. \textbf{D:} Waterfall plot of a single ASCII character from the `hello' sequence (`H') in its constituent bits. Note that the levitated hexagon used in \textbf{B} is a diferent one than that used for \textbf{C} and \textbf{D}.}
    \label{Fig:sketch_conceptual}
    \label{Fig:hello_world}
\end{figure}

\section{Experimental setup}

A conceptual schematic overview of the experiment is given in Figure \ref{Fig:sketch_conceptual}, along with the major results, while the detailed experimental setup is given in Figure \ref{Fig:experimental_layout}. The trap is formed by counterpropagating \SI{1560}{\nm} trapping light with a \SI{12.5}{\um} focus. We optically trap \SI{5}{\percent} doped \SI{5}{\um} diameter, \SI{200}{\nm} thick NaYF hexagons as shown in the SEM Appendix Figure \ref{Fig:SEM}. The pump light is provided by a separate laser (a pure photonics PPCL), which is fiber-coupled into the trapping region via a Thorlabs $2 \times 2$ PM 90:10 splitter -  since it is very close to the wavelength of the primary trapping laser. A wavelength division multiplexer (WDM) is used to counterbalance the insertion loss on the other side of the trapping region - but is not actively used in the experiment - for visual clarity the WDM is omitted from the experimental diagram. In contrast to the counterpropagating trapping light, the pump light only enters the trapping region from one side of the optical trap - this is specifically to avoid the creation of a second competing optical lattice. In particular,
pump light entering the trapping region from one side generates the potential systematic error of lowering the center of mass frequencies with increased optical power of the atomic resonance (since it pushes the test mass further away from the optimal/highest frequency trapping region), whereas a second competing lattice might raise or lower the center of mass frequencies depending on the alignment of the first and second lattice. Given the lattice overlap depends on wavelength, this could represent an unacceptable source of systematic error. 

The optical detection scheme for the levitated test masses is broadly explained in Ref.\cite{winstone2022optical}, with a few small changes having been made between \cite{winstone2022optical} and this work. In particular, the setup described in Ref. \cite{winstone2022optical} contains a detection channel called 'fiber-line', and in this work we also create an additional similar detection channel, in which a moveable free space detector is placed after the fiber output coupler of the experiment for an additional degree of tunability. All of the data in this work is taken with this new detection scheme. 

% the data taken is by the detector labeled as `fiber line to free space hyrbid' in the previous paper - in which trapping light propagated through the trapping region is re-coupled into the fiber system and then transmitted to a \textcolor{blue}{detector}. The one important change over previous work is that the `fiber-line' detector is displaced in free space to allow for the simultaneous detection of all modes.}

\section{Experimental Results}

\subsection{Shifting the central rotational frequency by changing the color of the applied pump light }
\label{section-momentum-mechanism-discussion}

\begin{figure}
    \centering
    \includegraphics[width=0.44\textwidth]{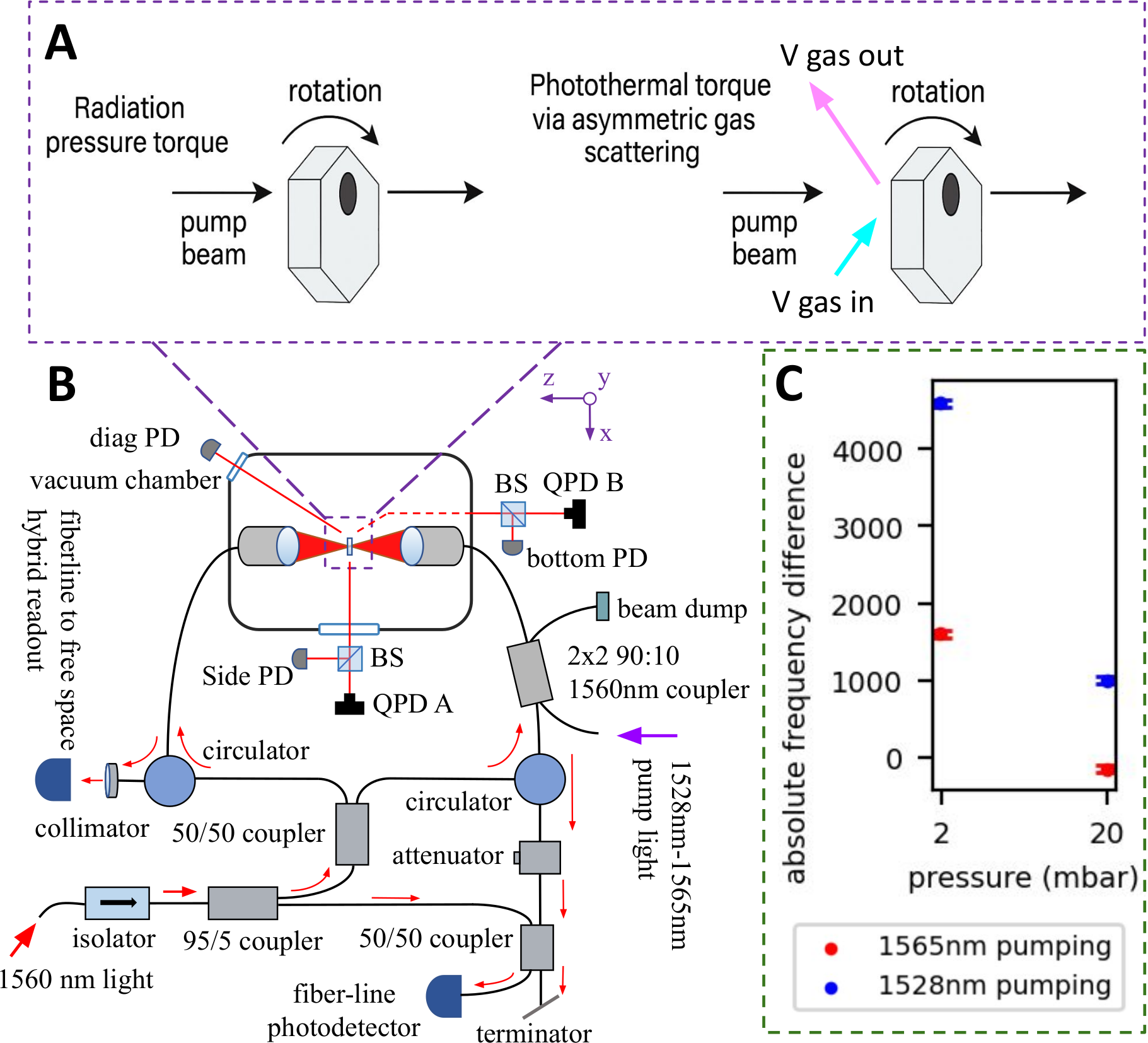}
    \caption{\textbf{A}: Angular momentum injection mechanisms. (left) In the first mechanism the momentum of injected blue-detuned photons is turned directly into rotational angular momentum by the photons being absorbed by the erbium ions embedded in the NaYF crystal matrix - the photons literally push the optically levitated hexagon to spin faster. (right) In the second mechanism, the blue-detuned photons are turned into rotational angular momentum by being absorbed by the erbium ions embedded in the NaYF crystal matrix as heat. The hotter surface temperature of the levitated hexagon then pushes incident background gas particles away faster than they impacted (the hot surface rocket effect), thus leading to a faster total free body spin rate. \textbf{B}: Experimental setup:\SI{0.5}{\W} of \SI{1560}{\nm} counterpropagating trapping light (provided by NKT C15 laser) is split in fiber and fed to the trapping site at a beam waist of \SI{12.5}{\um}, red detuned (\SI{1565}{\nm}) and blue detuned (\SI{1528}{\nm}) light (provided by a separate Pure Photonics PPCL laser) is fed to the trapping site along one arm of the counterpropagating fibers and thus trapping light. The power of the pump beam incident on the levitated hexagon can be varied across a range of \SI{0.1}{mW} to \SI{50}{mW}. \textbf{C}: The effect of the detuned photons on the spin rate as a function of pressure is explored experimentally over one order of magnitude, for two different wavelengths. For each pressure and wavelength, the frequency difference from \SI{31}{mW} to \SI{0.1}{mW} of pump power is shown.}
    \label{Fig:momentum_mechanisms}
    \label{Fig:experimental_layout}
    \label{Fig:experimental_setup}
\end{figure}

The free body rotational dynamics of optically levitated Rayleigh scale objects have been well studied in the past \cite{rashid2018precession}.
Furthermore, control over the rotational degrees of freedom has been shown previously by modulating the optical trapping power and pressure of the system
%\cite{bang2020five}\cite{kamba2023nanoscale}\cite{pontin2023simultaneous}
\cite{bang2020five, kamba2023nanoscale, pontin2023simultaneous}. Here, for the first time, we show control over the rotational degrees of freedom by modulating the wavelength of the incoming laser light.
Since the NaYF hexagons are doped with erbium ions (\SI{5}{\percent} by mass), the strongly wavelength dependent absorption spectrum of the erbium ion increases the net extinction cross section of the levitated object substantially at a certain wavelength, this enables this level of control over the free body dynamics with a relatively small amount of wavelength modulation.
%%%% maybe move this to the experimental description section
By controlling the wavelength of the pump laser, we
can change the center of mass rotation frequency of the levitated hexagon.

The damping experienced by the free body rotational motion of the hexagon is determined by the mostly constant background gas pressure inside the chamber. Increasing the torque being injected  into the system increases the speed at which the hexagon rotates, while decreasing it reduces the speed at which it rotates. %{AG: I re-ordered these paragraphs...}
We propose two mechanisms by which the pump light may be transformed into torque (see Figure \ref{Fig:experimental_layout}A). In the first, blue detuned photons are directly transformed into angular momentum through being absorbed at a higher efficiency by the erbium dopants that are on resonance with the wavelength of the light, if the center of the pump beam is slightly off axis with the center of the levitated hexagon.  

In the second mechanism blue detuned photons are indirectly transformed into torque by being absorbed at a higher efficiency by the erbium dopants and transformed into heat, this changes the internal temperature of the levitated NaYF prism: A gas molecule incident on a hot surface rebounds with a higher velocity characteristic of the higher surface temperature. Since momentum is conserved this leads to a thrust on the surface. This is often referred to as the Crookes radiometer effect \cite{keith2010photophoretic}. Naturally, this requires an inhomogeneous distribution of temperature over the levitated hexagon - a slightly asymmetric optical trap or distribution of dopants would be sufficient. Such localised heating effects causing localised surface thrust have been observed in levitated microparticles before \cite{millen2014nanoscale}\cite{ranjit2015attonewton}.

The pressure dependence of the rotation rate is shown in Figure \ref{Fig:experimental_layout}C over an order of magnitude in pressure. The difference in rotation rate between red and blue detuning is a function of pressure, indicating that gas collisions play a significant role in the observed dynamics.

To roughly estimate the torque involved in sustaining such a rotation process, we approximate the hexagonal prism as a cylindrical disc of radius $r=1.5$ $\mu$m, and thickness $t=0.2$ $\mu$m. Given the moment inertia for the ``coin-flip'' mode of $I=\frac{1}{4} \pi \rho r^4 t$, an angular acceleration $\alpha$ applied to the system in the presence of the viscous drag from the surrounding gas with a rotational damping rate of $\Gamma_{rot} = P \pi r^4/(I v_{gas}\sqrt{2\pi})[1 + \frac{\pi}{4} + \frac{t}{r} + \frac{t^2}{2r^2} + \frac{t^3}{4r^3}(1 + \frac{\pi}{6})]$ \cite{LISAdrag} would result in a steady state rotational velocity $\dot{\theta}_z = \alpha/\Gamma_{rot}$. Here $P$ is the background gas pressure and $v_{gas} \equiv \sqrt{k_B T/m_{gas}}$ is the characteristic thermal velocity of gas molecules of mass $m_{gas}$ at temperature $T$. To sustain a rotation speed of $\dot{\theta}_z \approx 4$ kHz, corresponding to the strongest observed rotational signal in the photodetector at twice this frequency $2\dot{\theta}_z \approx 8$ kHz, the needed torque is a approximately $9$ aN$\cdot$m. Such a torque could be imparted for example if the two optical beams are offset by a small amount $\delta$. Assuming an effective extinction power $P_{ext}$ that could be a combination of scattering and absorption, the angular acceleration applied to the system in this case is approximately $\alpha \approx P_{ext}\delta/(cI)$ where $c$ is the speed of light. We find that an offset of $\delta = 1$ $\mu$m and $P_{ext} = 30$ mW would be sufficient to provide the required torque, or an equivalent torque from radiometric forces would produce similar behavior. A more detailed estimate of such radiometric torques would require modeling the distribution of dopants, estimating the internal heating, and modeling the gas flow dynamics and is reserved for future work.

Here we have neglected the restoring force of the trap which would oppose free rotation of the hexagon for sufficiently small applied torques if the hexagon were starting trapped in its upright configuration as in Ref. \cite{winstone2022optical}. We expect these forces to be strongly acting only when the tilt $\theta_z$ of the hexagon is smaller than $\sim \lambda/8r$. For an axial trapping frequecy of $5$ kHz, We roughly estimate that external torques above a threshold value $\tau_{crit} \approx 2.5$ aN$\cdot$m would be sufficient to initiate the coin-flip rotation.

\begin{figure}
    \centering
    \includegraphics[width=0.44\textwidth]{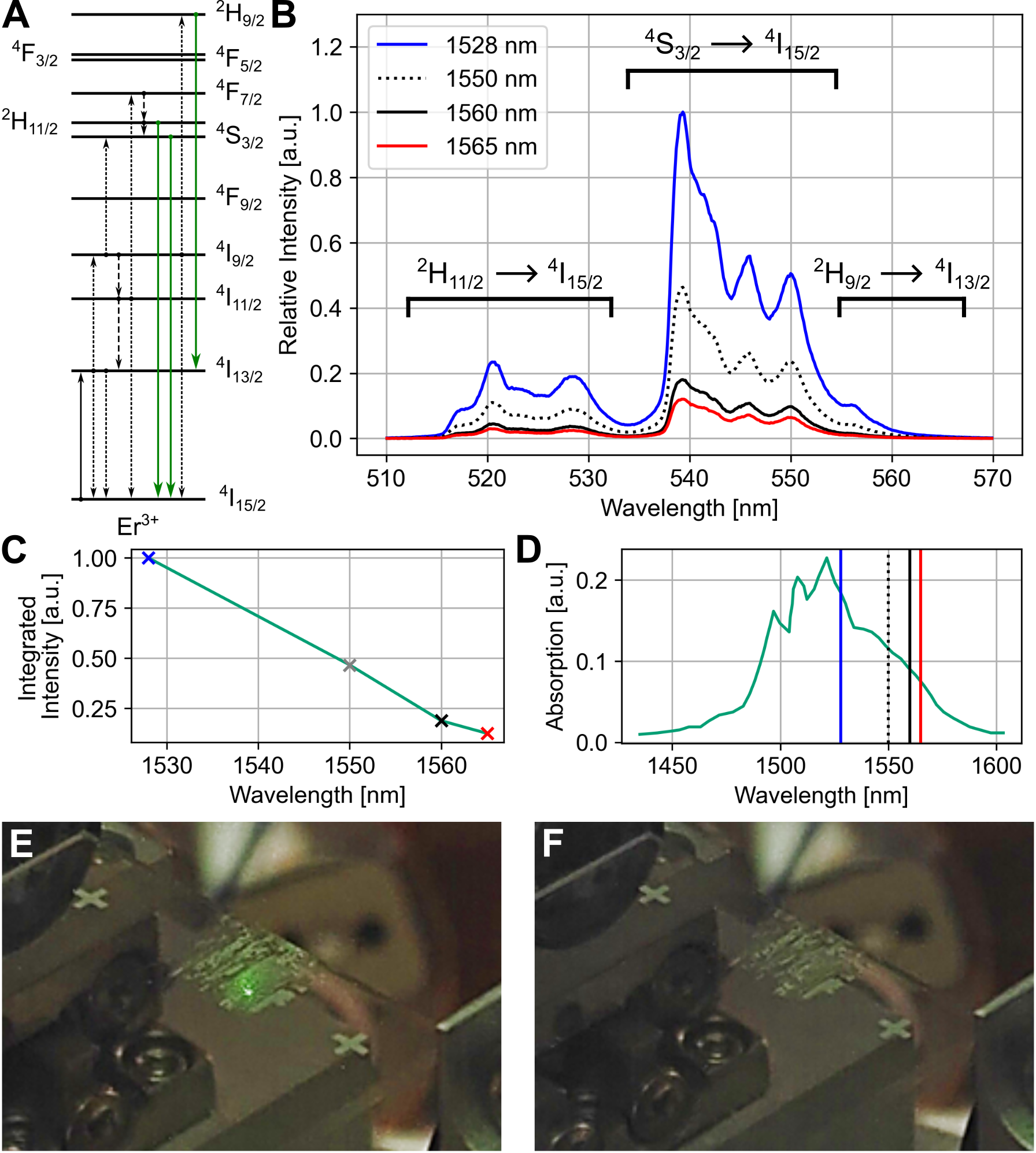} %F4_3.png
    \caption{
    % \textbf{A}: Erbium energy diagram and green emissions. \textbf{B}: Green erbium photoluminescence measured at different pump wavelengths. 
    % \textbf{C}: Optically levitated in free space just below the slide, \SI{5}{\percent} erbium doped \ch{\textbeta-NaYF} illuminated with \SI{1560}{\nano\meter} trap light emitting green light with sufficient intensity to image with a long exposure ($\sim$\SI{60}{\s}). \textbf{D}: Undoped optically levitated \ch{\textbeta-NaYF} suspended in free space immediately below the launch slide. Contrast and brightness of the images is digitally enhanced.
    %
    \textbf{A}: Erbium energy diagram and green emissions. \textbf{B}: Green erbium photoluminescence measured at different pump wavelengths. \textbf{C} Integrated intensity of the green erbium emission plotted against the pump wavelength.
    \textbf{D} Digitized and replotted literature absorption spectrum of \SI{5}{\percent} erbium doped \ch{\textbeta-NaYF} \cite{Ivaturi2013}. \textbf{E}: Optically levitated in free space just below the slide, \SI{5}{\percent} erbium doped \ch{\textbeta-NaYF} illuminated with \SI{1560}{\nano\meter} trap light emitting green light with sufficient intensity to image with a long exposure ($\sim$\SI{60}{\s}). \textbf{F}: Undoped optically levitated \ch{\textbeta-NaYF} suspended in free space immediately below the launch slide. Contrast and brightness of the images is digitally enhanced.
    }
    \label{Fig:fig3_greenglow}
\end{figure}

An image of strong green emission from upconverted photons is given in Figure \ref{Fig:fig3_greenglow} for a trapped particle. 
% The theoretical Erbium absorption spectrum is given in figure \ref{Fig:fig3_greenglow}, where our pumping wavelengths are also delineated, the simplified erbium level structure is given in figure \ref{Fig:fig3_greenglow} (top left), 
A simplified energy level diagram of the erbium dopants and pump wavelength-dependent photoluminescence is given in Figure \ref{Fig:fig3_greenglow} and the proposed multi photon relaxation pathway leading to green light emission is highlighted. The intensity of emissions in the top right inset follows absorption measurements from the literature \cite{Ivaturi2013} and illustrates the efficiencies at detuned wavelengths. As further explained in the Appendix the rotational motion visible in this experiment displays a significantly higher detection efficiency / signal to noise ratio (SNR) than the translational degrees of freedom. We attribute this to the larger amount of material being moved through a larger space in a free rotation compared to a translation. A free rotation of a high aspect ratio object with a diameter of microns moves its whole mass a full $2\pi$ steradians about its own axis. This is several orders of magnitude more matter moved through space than an objects translating some 10's of nanometers (thereby perturbing the incident light beam more). \footnote{Recent work has computed the information radiation patterns (IRPs) for bound librational states of motion through Fisher information approaches \cite{laing2024optimal}, however unbound free rotations were not considered in that work.}

% To investigate the comparative efficiency of the two effects we investigate the effect of detuning the pump light from blue detuned (with respect to the Erbium line) to red detuned (1528nm to 1565nm) over one order of magnitude in pressure. The difference in photon to angular momentum conversion efficiency suggests both mechanisms are simultaneously operational.  

%   this was pasted from section 4 and will be gone soon ---aboe

% \begin{figure}
%     \centering
%     \includegraphics[width=0.44\textwidth]{experimental layout.pdf}
%     \caption{Experimental setup. 0.5W of 1560nm counterpropagating trapping light (provided by and NKT C15 laser) is split in fiber and fed to the trapping site at a beam waist of 12.5um, red detuned (1565nm) and blue detuned (1528nm) light (provided by a separate pure photonics PPCL laser) is fed to the trapping site along one arm of the counterpropagating fibers and thus trapping light.}
%     \label{Fig:experimental_setup}
% \end{figure}

\subsection{Binary encoding of a 40 bit message onto the rotational degree of freedom via wavelength modulation - "Hello world"}

Rotational degrees of freedom are extremely sensitive to their boundary conditions and environment, variables such as; pressure, laser power, laser relative intensity noise (RIN), etc. Long term studies of rotating levitated micro and nanoparticles often observe significant variations in the rotational frequency over time - especially under non UHV conditions. Rotating particles are sometimes locked to an external reference to address this, leading to ultra stable mechanical rotors \cite{kuhn2017optically}.

To demonstrate our degree of substantial control over the system, over the level of the natural systematic errors of the instrument response and experiment noise, we chose to encode a binary message into the wavelength pump into the system, and then we read that message out in the center of mass rotational frequency of the levitated hexagon.

A traditional first case message to encode is often `hello world'. In this case, we chose to just encode the word `hello'. The message was encoded into a binary sequence by changing the wavelength of the applied pump beam. 
%Blue detuned (with respect to the Erbium resonant frequency) pump light at (\SI{1528}{\nm}) pump represents a bit of 1, Red detuned pump (\SI{1565}{\nm}) represents a bit of 0. The non pump light runs at an optical trapping power of \SI{500}{\mW} in a counterpropagating configuration (figure \ref{Fig:experimental_setup}). 
Blue detuned (with respect to the trapping beam) pump light (\SI{1528}{\nm}) represents a bit of 1, Red detuned pump light(\SI{1565}{\nm}) represents a bit of 0. The non pump light (\SI{1560}{\nm}) runs at an optical trapping power of \SI{500}{\mW} in a counterpropagating configuration (Figure \ref{Fig:experimental_setup}).
The encoded ASCII sequence is sufficiently resilient to long term/systematic drifts in the rotational frequency to encode a 40-bit message in wavelength space, which is then read out in the mechanical motion of the particle as the frequency of the rotational peak. This is shown in Figure \ref{Fig:hello_world}. Throughout the experiment either \SI{0.15}{mW} (low-power case) or \SI{31}{mW} (high-power case) of pump laser power reaches the levitated hexagon. The power dependence is explored briefly in Figure \ref{Fig:hello_world}B and Figure \ref{Fig:experimental_setup}C. At high pressure red-detuned pumping slightly reduces the spin speed of the hexagon with increasing power, while blue-detuned pumping increases the spin speed. At lower pressure increasing the pump power at both red- and blue-detuned pumping increases the spin speed.

\subsection{Long term scans - tennis racket effect}

Since angular momentum is constantly being injected by the pump light into the rotating hexagon, it does not immediately spin down, and can rotate for days or weeks on end. This allows us to observe long term rotational dynamics of the optically levitated test mass. Note that in our experimental setup, hexagons that are not doped with erbium tend to slow down their rotation on the time scale of tens of minutes to hours - (Figure \ref{Fig:before_during_after}), and then eventually stop rotating entirely. We attribute this to lack of a constant driving force keeping the rotation going.

%%%% I removed this discussion since the actual comparison to Daves data, accounting for pressure makes very little sense - our spin down times are actually better despite a less flat object.

% The decay time of 0.5hrs, in comparison to larger spin-up and spin-down/damping times (16hrs) for rotating vaterite spheres\cite{monteiro2018optical} suggests the large surface area of the hexagonal plates leads to a larger damping coefficient. \textcolor{red}{[AG: what about the pressure in each case? Is the vaterite experiment done at a comparable pressure? The spin-down rate will go down substantially in higher vacuum]}

The Dzhanibekov effect, or tennis racket effect, describes how objects with three distinct axes of inertia behave during free rotation. While they rotate stably around their longest and shortest axes, rotation around the intermediate axis is unstable, causing the object to flip about this axis, exhibiting metastable rotations in each orientation. We observe slow transfers in rotational frequencies (and thus angular momentum) of the levitated hexagon interspersed with rapid flips. These slow transfer invert signs (direction) at periodic points in the long term data set. We interpret these long-term periodic flips to be similar physics to that of the tennis racket effect (Dzhanibekov effect), in which a rotating object with the right mass distribution periodically flips about its prime axis of rotation \cite{van2017tennis}. A popular and intuitive demonstration of this effect is spinning a spanner(wrench) in zero gravity on the international space station. This effect can occur for both a classical and quantum mechanical rotor \cite{ma2020quantum}.

\begin{figure}
    \centering
    \includegraphics[width=0.48\textwidth]{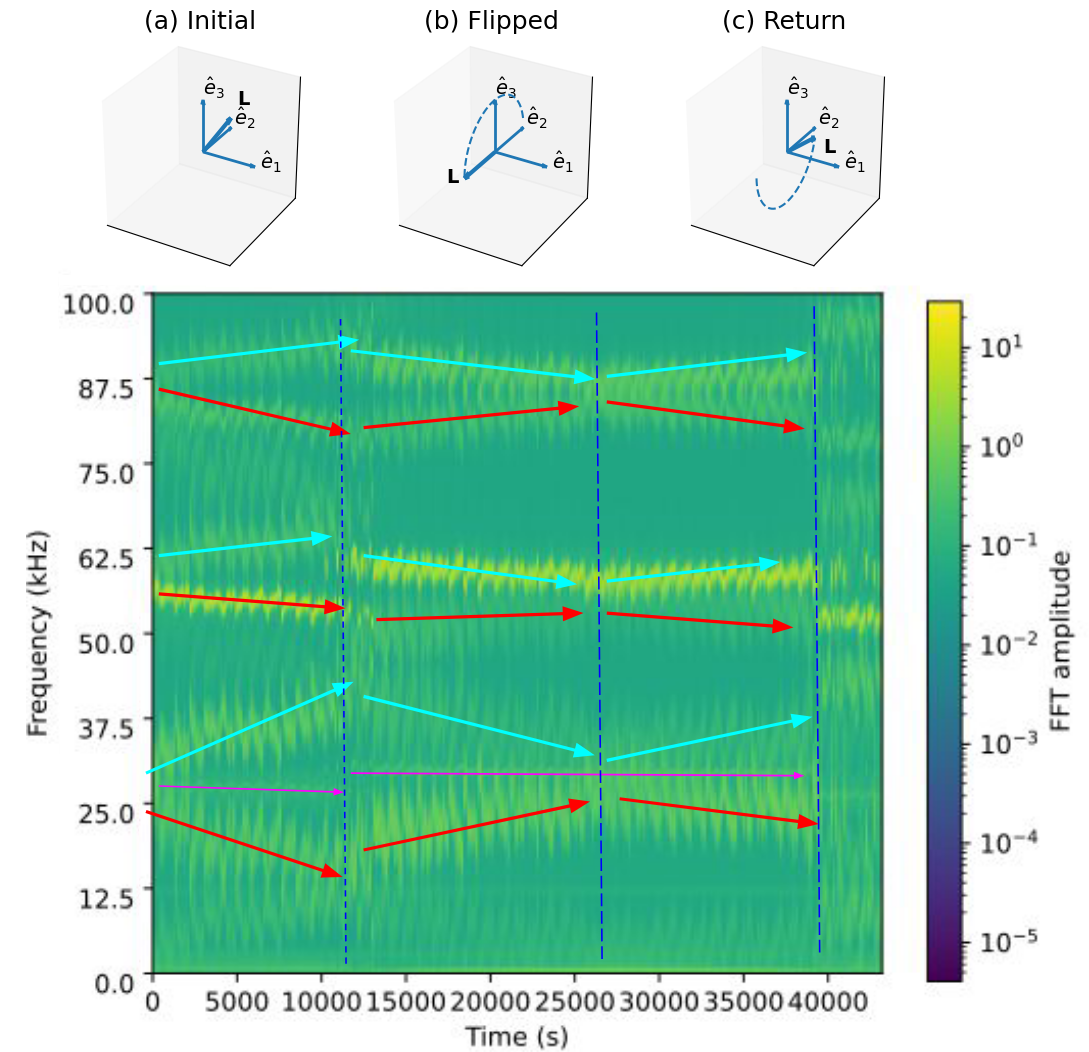}
    \caption{Tennis racket effect like flips delineated in a long term dataseries for a levitated mass. A conceptual example of a possible orientation of the hexagon during the experiment is presented.}
    \label{Fig:long_term_data_moreflips}
\end{figure}

Several features indicative of tennis racket like effects - strictly in the classical regime - are delineated in long term frequency space traces of a rapidly rotating optically levitated mass composed of NaYF hexagons. Two effects indicative of tennis-racket-like effects are shown in the dataset: 1) Slow drifts in the rotational frequencies leading up to 2) flipping/inversion points, figure \ref{Fig:long_term_data_moreflips}. Qualitatively this is the same dynamic behaviour as the flipping spanner/wrench freely floating on the ISS \cite{NASA_Dzhanibekov_effect_2015}.

% Finite element modeling of the hexagon - with best estimates of its mass distribution has enabled initial simulations attempting to derive the flipping period from experimental conditions. 

% \textcolor{red}{[AG: it sounds strange we state this but don't include results. should we say that in principle this could be done?]} 

% The parameter space of hexagon/hexagon agglomerates mass distributions that allow for flipping periods on the order of that observed is highly degenerate. On the one hand this makes an exact parameter match difficult, on the other it suggests that it is reasonable to identify this as tennis-racket-like physics. Since the primary scope of this work is experimental, an exhaustive parameter space search and the construction of a less naive model is left for future work

Finite element modeling to derive the exact flipping period of the hexagon from the set of physical inputs given in the work will be a topic of near term future numerical studies, and is beyond the scope of this experimental work.

However, it should be noted there are several pieces of physics that are not included in the simple naive tennis racket model, such as a net flow of energy into the system driving the rotational motion, significant damping, significant center of mass motion in non rotational degrees of freedom and the existence of non trivial couplings to additional degrees of freedom in the system. All of these are present in our experiment and are non-trivial to model.

% The naive tennis racket model, and the extensions used for our computational brute force parameter space search are derived in Appendix \ref{Tennis_racket_derivation}

\subsection{Green light generation - NIR to visible upconversion}

When optically trapped, the erbium-doped \ch{\textbeta-NaYF} emits green light upconverted from the \SI{1560}{nm} trapping laser. This single particle emission is bright enough to be imaged with a regular camera and a long exposure time (about \SI{60}{s}) as shown in Figure \ref{Fig:fig3_greenglow}. The photoluminescence (PL) is caused by an overlap of the trapping wavelength with the \textsuperscript{4}I\textsubscript{15/2} \textrightarrow \textsuperscript{4}I\textsubscript{13/2} electronic transition, resulting in absorption. Consecutive excited-state absorption (ESA), energy transfer (ET) between erbium dopants, and non-radiative relaxation will lead to a population of the \textsuperscript{2}H\textsubscript{11/2} and \textsuperscript{4}S\textsubscript{3/2} energy levels, which emit green photons when radiatively relaxing to the \textsuperscript{4}I\textsubscript{15/2} ground state as schematically illustrated in Figure \ref{Fig:fig3_greenglow}. Furthermore, the emission from the \textsuperscript{2}H\textsubscript{9/2} to the \textsuperscript{4}I\textsubscript{13/2} excited state falls into the green wavelength range as well \cite{Renero-Lecuna2011}. The independently obtained spectroscopic data in Figure \ref{Fig:fig3_greenglow} show emissions from transitions to the electronic ground state of erbium, but no contributions from excited state absorption. While the irradiance was much lower than in the trapping setup, this absence is promising for complication-free ratiometric thermometry \cite{Xia2021a}. The relative and integrated intensity of the green photoluminescence illustrates the influence of the pump wavelength with a decrease from blue-detuned to red-detuned excitation, showing the same trend as literature absorption measurements of \SI{5}{\percent} \ch{Er:\textbeta-NaYF} \cite{Ivaturi2013}. It should be noted that while the emission appears green to the eye, there are many more emission lines due to the rich electronic structure of \ch{Er^{3+}}. Spectra from \SI{500}{\nm} to \SI{1000}{\nm} are presented in Appendix Figure \ref{Fig:FullUpconversion}.

The observation of strong emission points towards two potential use cases in future work. First, the upconverted light could be used for virtually background-free detection of the particle dynamics in the trap. Second, in situ contactless thermometry is possible given that the ratio of emission from the \textsuperscript{2}H\textsubscript{11/2} and \textsuperscript{4}S\textsubscript{3/2} excited states is known to be very sensitive to changes in temperatures.

\section{Conclusion}

In summary, we demonstrate the first wavelength-detuning-based driven rotational control of a high aspect ratio levitated \textbeta-NaYF microhexagon doped with erbium. This represents the first foray into \textit{wavelength} based rotational control in an optically levitated system.

The ability to tune the free rotational frequency via wavelength modulation suggests a pathway to improved rotational stability and therefore sensitivity that does not depend exclusively on modulating power or polarization at the trap (as has been done previously in the literature). To achieve high levels of gyroscopic performance, systems based on micron scale optically levitated test masses require high degrees of feedback control \cite{zeng2024optically}, being able to tune several variables independently that are not highly coupled to each other is a valuable element to have for the operation of any closed loop system.

Finally, other optically levitated morphologies (such as spheres, dumbbells and nanorods) have been developed to routinely enable a high degree of control of rotational degrees of freedom. In developing control systems for high aspect ratio particles we note a few general use cases for the rotational degree of freedom for these particles: (i) The rapidly rotating optically levitated hexagons produce an optomechanical frequency comb of up to the 20th harmonic - opening up potential applications for broadband sensing with mechanical frequency combs; (ii) Given the quantity of light emitted in the green by upconversion, there exists significant potential for background and trapping light free detection by counting emitted green photons; (iii) due to their extremely high aspect ratio of optically levitated disk likes \cite{winstone2022optical}, they should make excellent pressure sensors while rotating - with a high mass to area ratio in the parachute-like axis and an extremely low ratio in the anti-parachute axis - potentially allowing for shape dependent calibration within the same sample. Furthermore the fine control over temperature and spin rate may present applications for any rotation based optofluidic propulsion, mixing, or sensing\cite{ortiz2021laser}. These general use cases aside, a potentially highly impactful application of our work here is likely gyroscope development, inertial sensing, and navigation. With this work we have expanded the toolbox of possible mechanisms to control rotation in optically levitated spinning mass gyroscopes.

Lastly and finally we observe long-term rotational dynamics qualitatively consistent with tennis-racket-like behavior in an optically levitated system - also for the first time to the best of the authors knowledge. This is expected to have potentially significant applications in high mass matter wave interferometry due to the potential natural resistance of these modes to decoherence \cite{ma2020quantum}.

\clearpage
\section{Appendix}

\subsection{Comparison: spinning and non tennis racketing, spinning and tennis racketing, and non spinning hexagons.}

\begin{figure}
    \centering
    \includegraphics[width=0.99\linewidth]{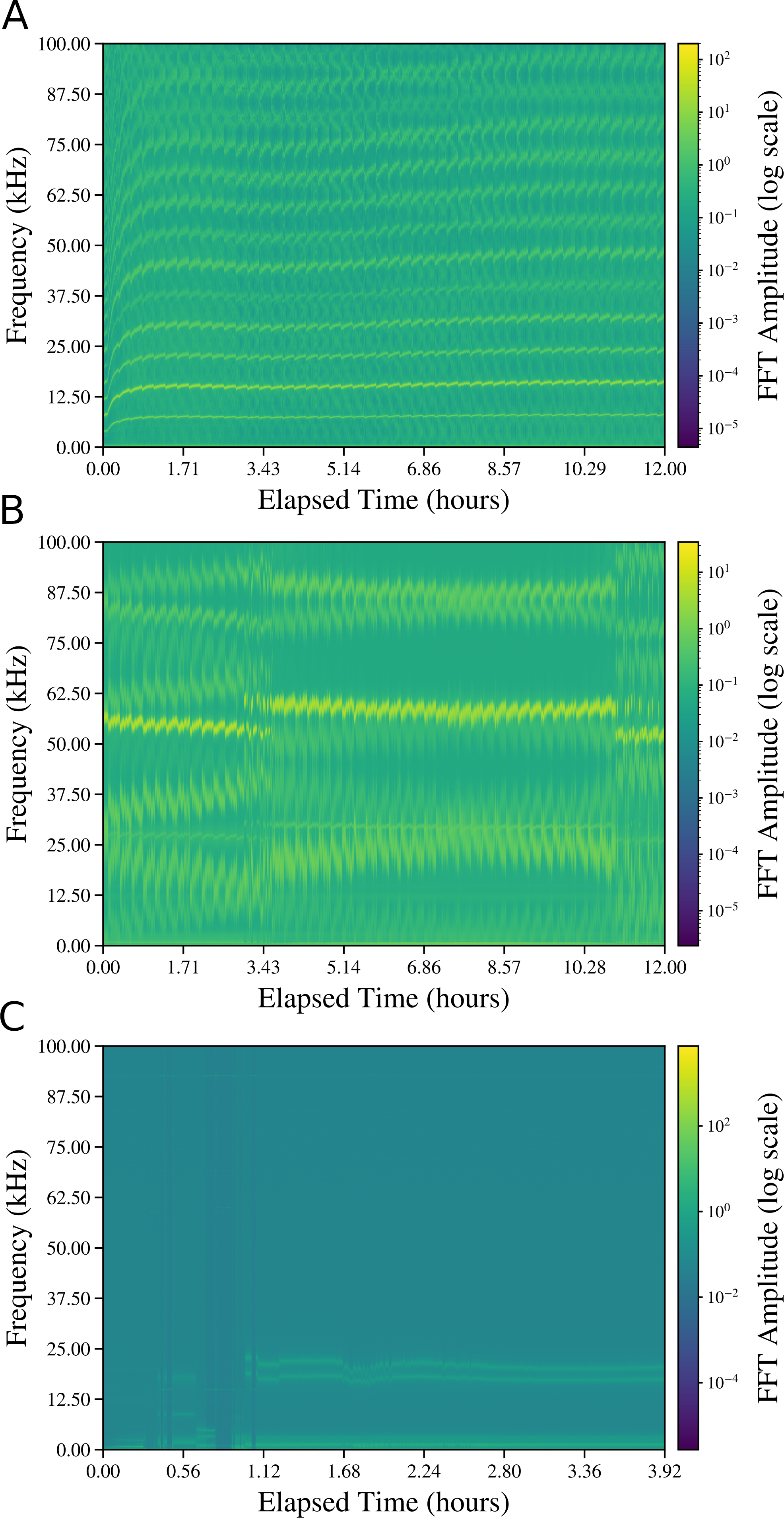}
    \caption{Comparison: A: spinning and non tennis racketing levitated hexagon, B: spinning and tennis racketing levitated hexagon, C: non spinning levitated hexagon translating in bound vibrations/librations in the standing wave. The vertical interruptions in C are due to intermittent failures in the data system over long time scales and do not represent particle loss.}
    \label{fig:tennis_spinning_translating_datacomparison}
\end{figure}

In figure \ref{fig:tennis_spinning_translating_datacomparison} data is presented for optically levitated hexagons under several district states. A:  a spinning and non tennis racketing levitated hexagon undergoing free rotations, B: a spinning and tennis racketing levitated hexagon undergoing free rotations and experiencing the directional flips consistent with a tennis racket like effect. C: a non spinning levitated hexagon translating in bound vibrations/librations in the standing wave potential defined by the counter propagating beams. The vertical interruptions in C are due to intermittent failures in the data system over long time scales and do not represent particle loss. Experimental conditions were a trapping power of 500mw and a roughing pump pressure of 2mbar.

\subsubsection{Effect of the vacuum pump on the rotational frequency}

Since for a rotating system the free body rotating frequency is directly dependent on the systems pressure, the pressure set point of the vacuum pumping system and subsequent oscillatory behavior around said setpoint manifests as a distinct oscillation of the free-body rotational frequency in the long term datasets of rotating hexagons, this frequency is significantly higher than the other identified dynamics within the system. Note the approx 15 minute oscillations in figure \ref{fig:tennis_spinning_translating_datacomparison} subplots A and B.

\subsection{Auxiliary data - full recorded frequency spectrum of the FFT for the red/blue pumping experiment.}

\begin{figure}
    \centering
    \includegraphics[width=0.48\textwidth]{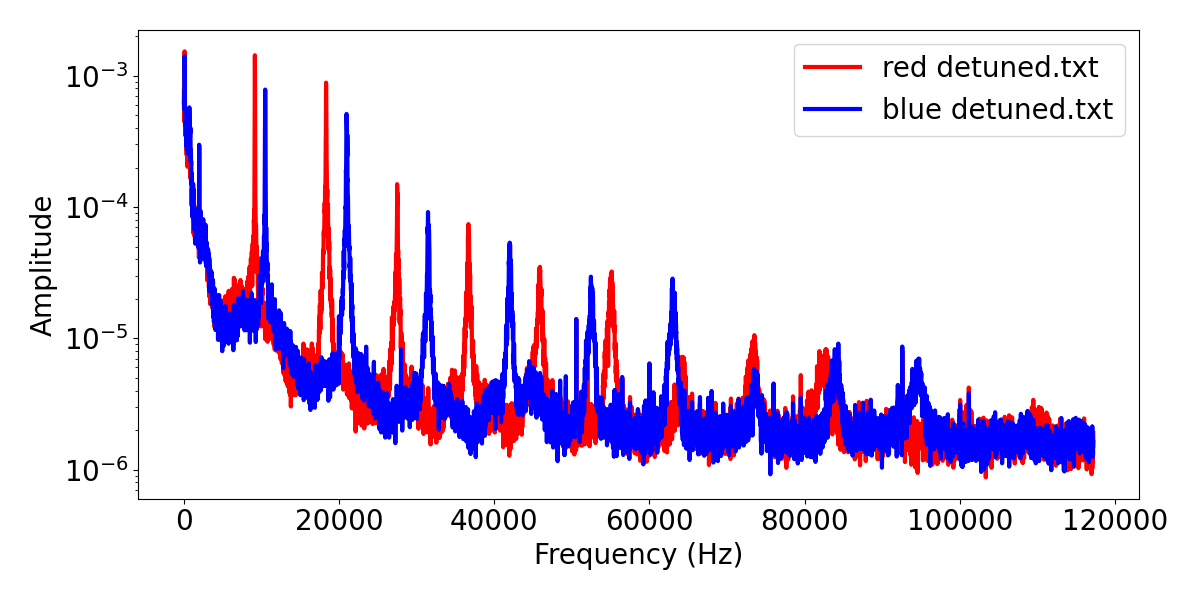}
    \caption{Full frequency spectrum of the rotating hexagon shown initially in Figure \ref{Fig:hello_world}, the figure depicts the red and blue detuning of the free rotational motion of an optically levitated hexagon when being hit with red and blue detuned light. The applied pump power is \SI{31}{mW} in both cases. }
    \label{Fig:fig1zoomedout}
\end{figure}

For completeness, Figure \ref{Fig:fig1zoomedout} shows the full frequency spectrum of the data given initially in Figure 1, showing all the higher order harmonics of the freely rotating mode up to the $\sim 20$th order. Of note is that the rotational mode detection efficiency completely overpowers that of the translational mode detection efficiency. When a high-aspect-ratio levitated hexagon starts spinning, e.g. in its ``coin-flip'' mode, its translational dynamics tend to be difficult to observe in the motional power spectrum.%AG rephrased this --no longer be visible.

% \subsection{Detection of a freely rotating massive object}

We primarily attribute this difference in detection efficiencies to the amount of material that moves through space - associated with a particular degree of freedom. A large change in the intensity of light striking the photo-detector occurs as the light reflecting from the face of the hexagon sweeps out a very large set of angles as the particle rotates.

For means of comparison, the translational thermal motion (assuming it is simple harmonic oscillator in contact with a thermal bath at $\sim$\SI{300}{\kelvin}) for a ~$5$ $\mu$m diameter $200$ nm thick levitated hexagon is on the order of magnitude of tens of nanometers \cite{winstone2022optical}. By contrast the freely unbound rotating motion of a disc moves its entire mass about its own axis, at a radius diameter of~\SI{5}{\micro\meter}.

In the case of another commonly studied rotating optically levitated object - the nanodumbell \cite{rashid2018precession} the amount of material moving through space is about equal for the rotational degrees of freedom - the radius of the dumbbell is of order $\sim$ \SI{100}{\nano\meter}, with thermal motion on the hundreds of nanometers scale. We attribute this difference of characteristic scales (the translational and freely rotating modes of small low aspect ratio objects and larger high aspect ratio objects) as the reason for why the x, y, z modes disappear almost completely in the detectable output for the optically levitated hexagon while all modes remain visible in the case of rapidly spinning optically levitated dumbbells \cite{rashid2018precession}.

The scattered information patterns for a rotating / librating optically levitated test mass is a topic for future investigation, building on existing methods \cite{laing2024optimal}.

\subsection{Effect of blue and red light detuning on the optomechanical damping}

For the model of the linear translational degrees of freedom of an optically levitated harmonic oscillator, the damping tends to be dominated by collisions with background gas molecules.

The dependence of this damping upon the temperature of the background gas is an extremely complex function of the temperature itself, pressure, gas species, and the geometry of the object. However in the specific limit explored in this paper, we find that the damping tends to increase as we increase the amount of applied blue detuned pump light - which is likely increasing the internal temperature of the levitated hexagon.

\begin{figure}
    \centering
    \includegraphics[width=0.44\textwidth]{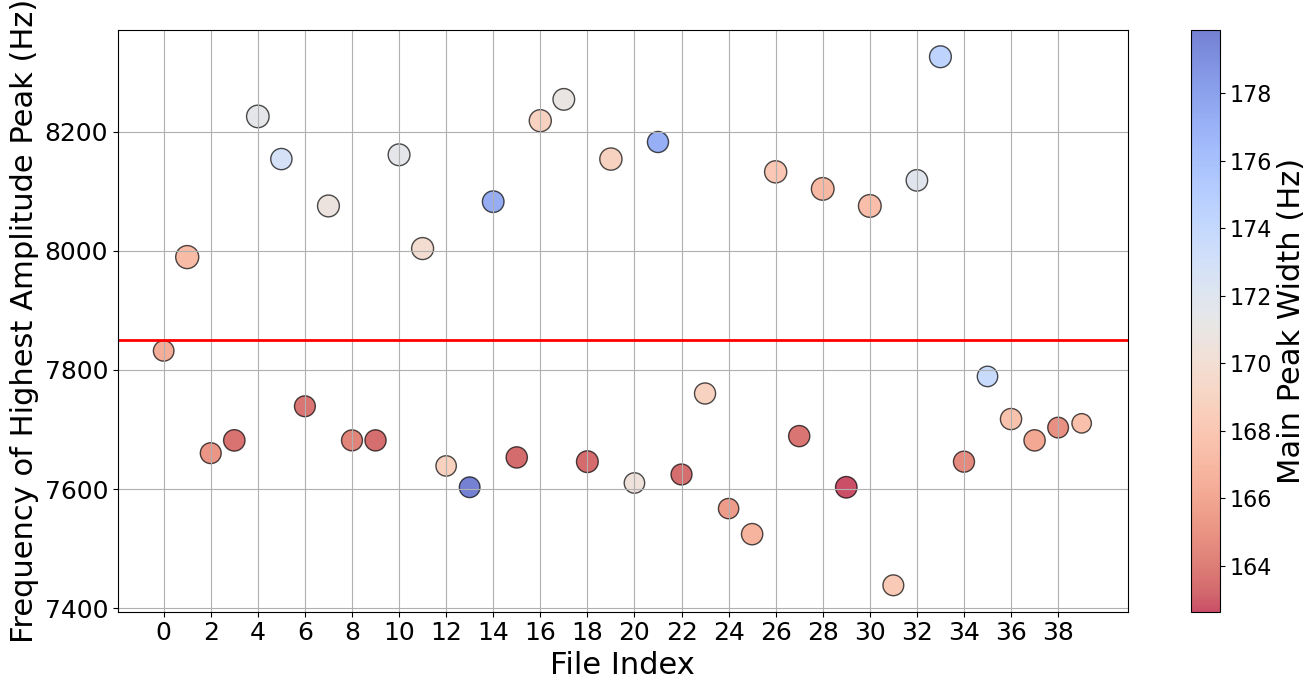}
    \caption{The same 'Hello World' dataset as in figure \ref{Fig:hello_world} is analyzed to investigate correlation between increased internal temperature of the levitated hexagon from being irradiated with blue detuned light and higher optomechanical damping. The frequency of the highest-amplitude spectral peak for each measurement file under blue and red detuned pumping conditions is plotted against the file index in the same order as in figure \ref{Fig:hello_world}. As in figure \ref{Fig:hello_world} - blue detuned datapoints have a higher frequency, while red detuned datapoints have a lower frequency - leading to the bimodal distribution in frequency space. Marker size is proportional to the peak height (amplitude). The colour of the points on a blue-red scale denotes the width of the measured peaks - corresponding to the optomechanical damping. On average, the peaks on the top half of the bimodal distribution - higher frequencies - blue detuned - tend to be wider in frequency space, this is consistent with increased damping due to internal heating of the levitated particle. (for peak detection and peakfinding, some smoothing of the dataset is required in frequency space, in this case we pick $\sigma$ = 64, representing the standard deviation of the Gaussian distribution convolved with the measurement signal)}
    \label{Fig:Thickerwithredpumping}
\end{figure}

The dataset shown in Figure \ref{Fig:Thickerwithredpumping} consists of 40 measurement files that already had Fourier transforms applied and was processed on the Northwestern Quest Supercomputing Cluster using the Quest-Analytics JupyterHub environment. Signal processing was performed using the \texttt{SciPy} library (specifically \texttt{scipy.ndimage} and \texttt{scipy.signal}), along with \texttt{NumPy} for numerical manipulation and \texttt{Matplotlib} for visualization. Peak analysis in particular made use of functions from \texttt{scipy.ndimage} and \texttt{scipy.signal}.

The voltage signal was first smoothed to reduce high-frequency noise using a one-dimensional Gaussian filter:
\[
V_{\mathrm{smooth}}(t) = \mathrm{gaussian\_filter1d}\!\left(V(t),\, \sigma\right).
\]
Several values of the Gaussian width parameter $\sigma$ were tested in order to identify an optimal balance between noise suppression and preservation of peak structure.

Peak detection was then carried out using \texttt{scipy.signal.find\_peaks}. To determine a consistent threshold for peak prominence across all signals, a data-driven prominence threshold was defined as

% \[
% \scalebox{0.85}{
%   \text{\[
% P_{\mathrm{threshold}} = \left[\mathrm{percentile}\!\left(V_{\mathrm{smooth}}, 95\right) - 
% \mathrm{percentile}\!\left(V_{\mathrm{smooth}}, 5\right)\right] \times 0.1
% \]}
% $}
% \]

{\small
\[
P_{\mathrm{threshold}} =
\left[\mathrm{percentile}\!\left(V_{\mathrm{smooth}}, 95\right) -
\mathrm{percentile}\!\left(V_{\mathrm{smooth}}, 5\right)\right] \times 0.1
\]
}

This percentile-based scaling ensured that peak selection was robust to variations in overall signal amplitude. Peaks exceeding this prominence threshold were retained, and their widths were subsequently extracted using the \texttt{peak\_widths} function for additional quantitative characterization
%\cite{Harris2020Array}\cite{Virtanen2020SciPy}\cite{Hunter2007Matplotlib}
\cite{Harris2020Array, Virtanen2020SciPy, Hunter2007Matplotlib}.

\subsection{Rotational dynamics - further observations}

\subsubsection{Single hexagons can be trapped 'normally', start spinning, and then return to being normally trapped.}

\begin{figure}
    \centering
    \includegraphics[width=0.49\textwidth]{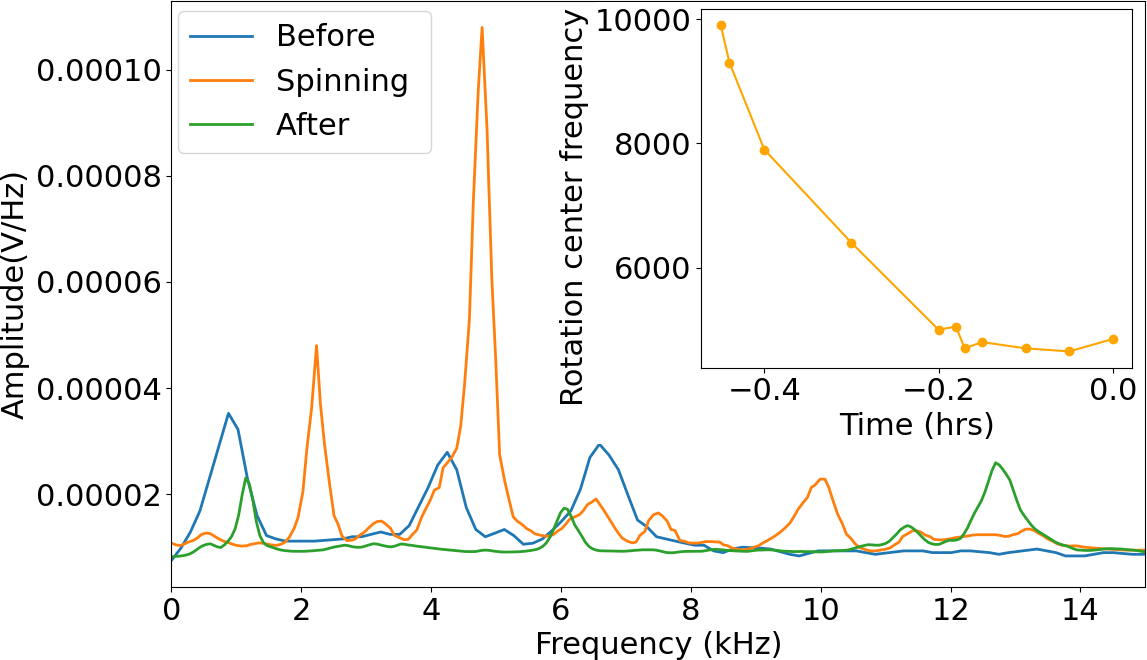}
    \caption{Main figure: In this dataset an undoped optically levitated NaYF hexagon is initially observed to be levitated normally, with all librational degrees of freedom bound within the potential of the standing wave and the flat face of the hexagon is aligned to the standing wave propagation axis. The hexagon is then perturbed by a random background fluctuation and begins to rotated freely about its axis, first slowing down to a lower rotational free body rate and then eventually returning to bound motion standing up within the standing wave optical potential. The transition spinning motion is accompanied by an increase in the signal to noise ratio / Q factor of the most dominant peak of almost an order of magnitude. The significance of this dataset is that 'single' well-formed hexagons can exhibit high speed rotation - 'dumbbell' like configurations are not required, although dumbbells (accumulations of two or more hexagons into bimodal mass distribution) may also be levitated and rotating within the trap in other runs. NaYF hexagons without erbium dopant tend to spin down in a matter of hours. Inset: The freely rotating motion of the optically levitated hexagon rings down without active injection of angular momentum via optical pumping of atomic dopants into momentum to maintain the rotation. The pressure is kept constant at \SI{2}{mbar} throughout the experiment.}
    \label{Fig:before_during_after}
    \label{Fig:ringdown}
    \label{fig:ringdown}
\end{figure}

Figure \ref{Fig:before_during_after} shows the \textbf{same} hexagon before starting a spin cycle, while rotating, and after spinning down into a well behaved bound position. This demonstrates that 'single' hexagons can display free body unbound rotations and that a dumbbell like shaped cluster is not required. This is a different hexagon than those used for the other results of this work, however it is trapped in the same setup under similar experimental conditions of power and pressure. The 'ringdown' of this hexagon during the spinning state is given in the inset of Figure \ref{Fig:ringdown}.

\subsubsection{Rotating and non rotating trapped hexagons comparative PSDs}

\begin{figure}
    \centering
    \includegraphics[width=1\linewidth]{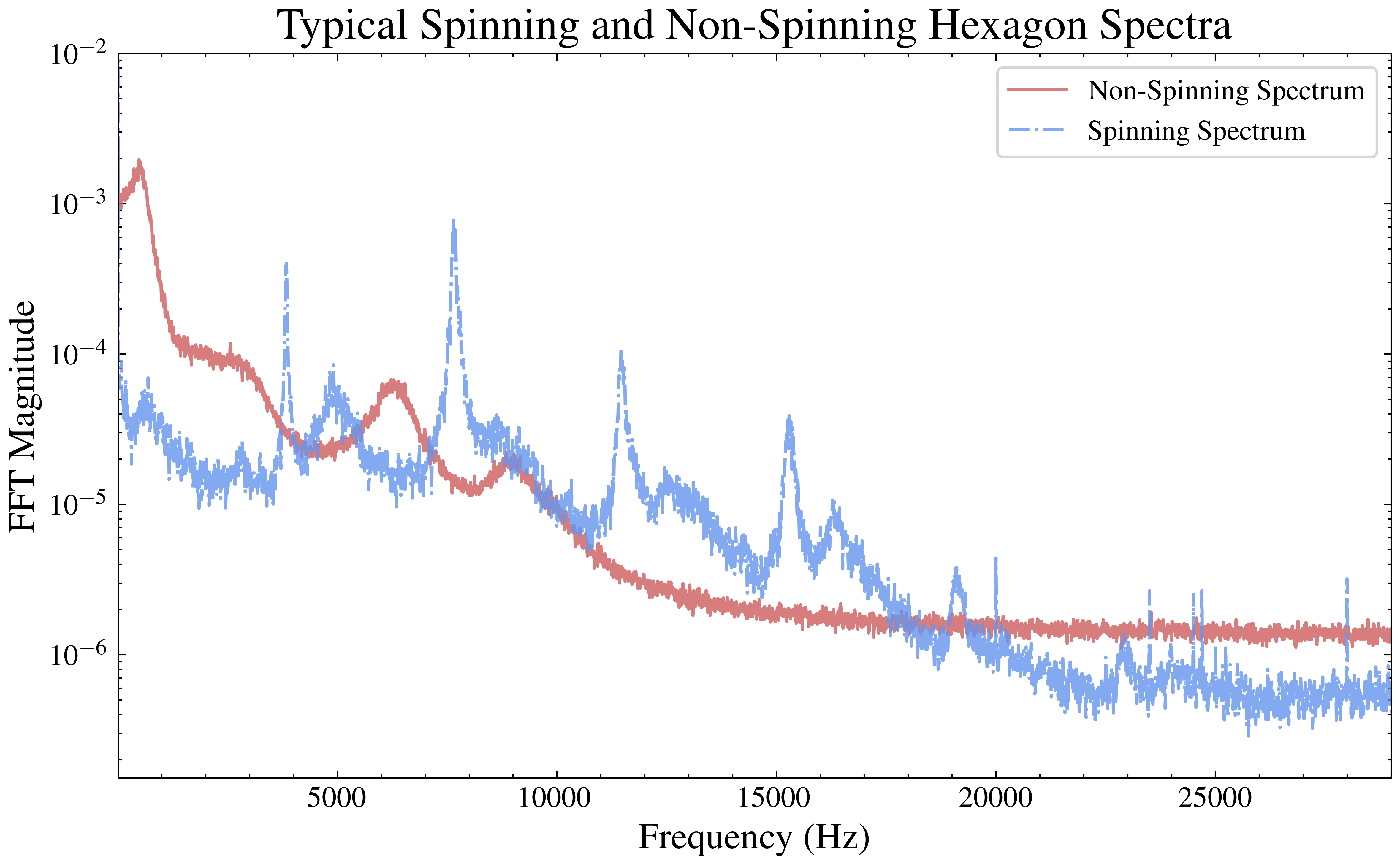}
    \caption{Comparison of hexagons bound to the optical lattice translating over small amplitudes and freely rotating hexagon PSDs overlaid.}
    \label{fig:bound_and_translating_comparison}
\end{figure}

Figure \ref{fig:bound_and_translating_comparison} shows two different hexagons, one bound in stable translations and librations, and one undergoing free rotations about its axis - taken under similar experimental conditions. The PSDs of freely rotating high aspect ratio hexagons tend to have one dominant frequency with many harmonics - the translational frequencies are almost entirely suppressed out of the PSD. Non freely rotating hexagons display the usual $\leq$ degrees of freedom. It should be noted however that identification of an optically levitated objects shape and dynamics from a single PSD is extremely difficult since the parameter space is so degenerate. The greatest point of identification for rotational dynamics is the rotational frequency dependence on pressure. 

\subsubsection{Rotating hexagons with no Erbium dopants or active pumping slowly spin down on their own}

In figure \ref{Fig:ringdown}, an optically trapped hexagon at a constant pressure with no atomic dopants gradually loses all its angular momentum to the surrounding gas molecules over a period of 0.5hrs, the rotational frequency spins down from 10KHz to about 5KHz before resuming a stable trapping configuration in the standing wave.

\subsection{Particle Synthesis and Characterization}

\begin{figure}
    \centering
    \includegraphics[width=0.44\textwidth]{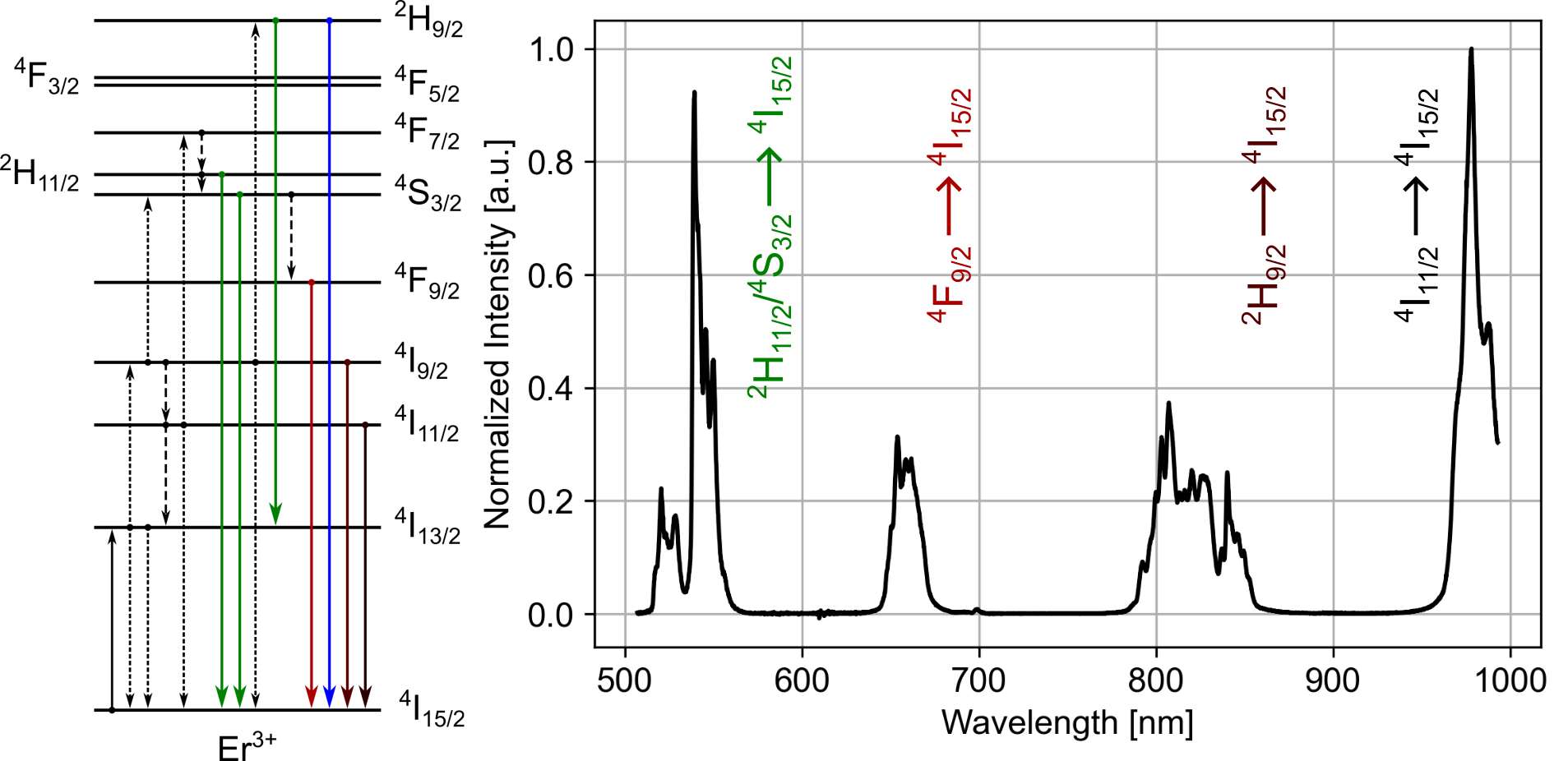}
    \caption{Energy level diagram of erbium and measured upconversion of \ch{Er:\textbeta-NaYF} under \SI{1560}{\nm} illumination.}
    \label{Fig:FullUpconversion}
\end{figure}

\begin{figure}
    \centering
    \includegraphics[width=0.24\textwidth]{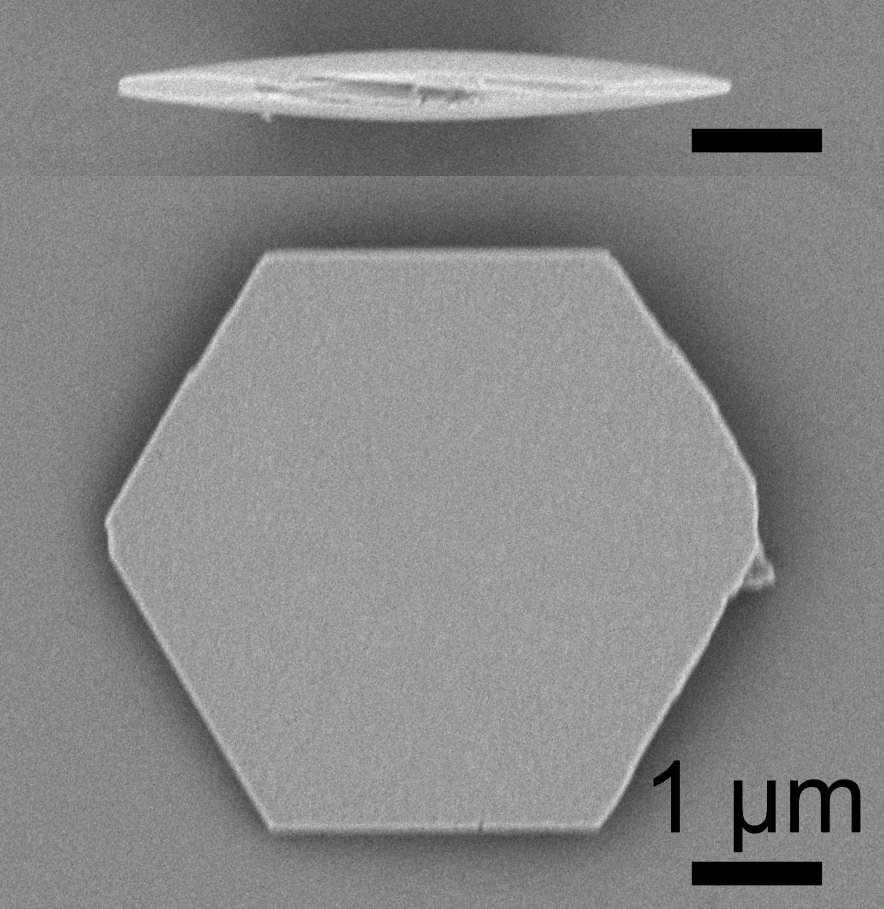}
    \caption{SEM images of synthesized \SI{5}{\percent} erbium doped \ch{\textbeta-NaYF} microcrystals. Note the slightly non regular geometry of the hexagon - this will lead to a slight mass asymmetry around all possible axes of rotation.}
    \label{Fig:SEM}
\end{figure}

\begin{figure}
    \centering
    \includegraphics[width=0.44\textwidth]{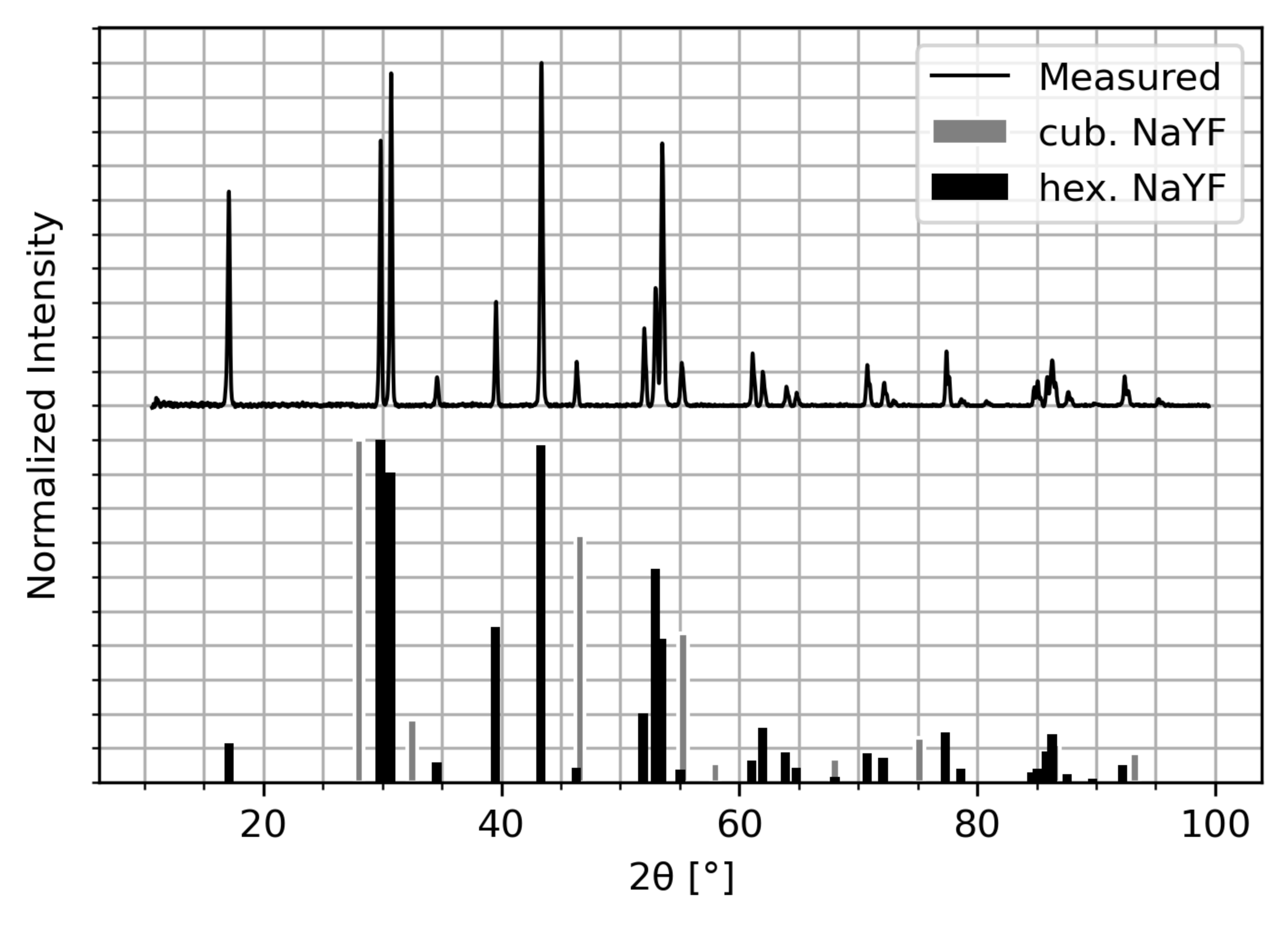}
    \caption{XRD pattern of synthesized \SI{5}{\percent} erbium doped \ch{\textbeta-NaYF} microcrystals. A comparison with database entries of cubic \ch{\textalpha-NaYF} (PDF 01-075-6676) and hexagonal \ch{\textbeta-NaYF} (01-092-0779) indicates phase purity.}
    \label{Fig:XRD}
\end{figure}

Undoped and \SI{5}{\percent} \ch{Er:\textbeta-NaYF} disks were synthesized following the hydrothermal procedure developed by Felsted et al.~with slight alterations \cite{petercrystalgrowth}. Yttrium(III) (\ch{YCl3 * 6 H2O}, $>$\SI{99.999}{\percent}, Sigma Aldrich) and for the doped case erbium(III) chloride hexahydrate \ch{ErCl3 * 6 H2O}, $>$\SI{99.9}{\percent}, Sigma Aldrich) were dissolved in ultrapure water (Barnstead GenPure) to obtain \SI{4}{\mL} of rare-earth solution containing a total of \SI{1.125}{\mmol}. Ethylenediamintetraacetic acid (\ch{EDTA}, $>$\SI{99.995}{\percent}, Sigma Aldrich) was dissolved in \ch{H2O} using \SI{5}{M} sodium hydroxide solution (obtained from $>$\SI{97}{\percent} \ch{NaOH}, Fisher Chemical) to nominally obtain \SIlist{5}{\mL} \ch{Na2EDTA} solution containing \SI{1.125}{\mmol}. 
The chelating solution was added to the rare-earth solution in a Teflon liner while stirring and stirred for \SI{10}{\min}. The fluoride precursor solution was obtained by dissolving \SI{4.5}{\mmol} (\ch{NaF}, $>$\SI{99.5}{\percent}, EMD Chemicals) in water to obtain \SI{5}{\mL}. After adding the fluoride solution to the chelated rare-earth solution under stirring and consecutive mixing for \SI{30}{\min} the Teflon liner was closed and placed in a hydrothermal autoclave (Parr Instruments 4747). The autoclave was transferred into a preheated oven (Thermo Scientific Heratherm) at \SI{220}{\degreeCelsius} and the synthesis allowed to proceed for \SI{72}{\hour}. The precipitate was collected by centrifugation after removing the autoclave from the hot oven and natural cooling to room temperature. The undoped or \SI{5}{\percent} \ch{Er:\textbeta-NaYF} was washed three times each using \SI{30}{\mL} of \ch{H2O} or ethanol (\SI{200}{proof}, Decon Labs.) and dried in an oven at roughly \SI{80}{\degreeCelsius}. To remove any remaining surface adsorbed \ch{EDTA} the dry powder was calcined in a tube furnace (Thermo Scientific Lindberg/Blue M) at \SI{300}{\degreeCelsius} with a \SI{30}{\min} ramp and \SI{4}{h} hold. 

Scanning electron microscopy images were obtained on an Apreo 2 (Thermo Fisher Scientific) using an acceleration voltage of \SI{2}{kV}, a current of \SI{13}{pA} and a backscatter detector.

Powder X-ray diffraction patterns are obtained using a Cu X-ray source operated at \SI{50}{\kV} and \SI{1000}{\uA} and a 2D detector (Pilatus 100K) on a Bruker D8 Discover.

Upconversion spectroscopy of the \SI{5}{\percent} \ch{Er:\textbeta-NaYF} was obtained by illuminating dry powder on a cover slip using a tunable telecom laser (Santec TSL-710) focused by a 50x objective (Mitutoyo Plan Apo NIR). The spot sizes are around \SI{5.5}{\um} resulting in irradiances from \SIrange{1.0}{1.1}{\kW \per \cm^2}. The photoluminescence was collected by a second 50x objective (Mitutoyo Plan Apo NIR) and measured using an EM-CCD (BLAZE 100-HR, Princeton Instruments) mounted on a spectrometer (HRS-500, Princeton Instruments) after passing a \SI{1000}{\nm} short-pass filter (FESH1000, Thorlabs).

Material characterization results are presented in Figures \ref{Fig:SEM} and \ref{Fig:XRD}. The mean particle diameter and height were \SI{4.9(1.0)}{\um} and \SI{210(60)}{\nm}. The XRD pattern indicates phase pure \ch{\textbeta-NaYF}.

% Figure \ref{Fig:Tennis racket flip animation} shows a conceptualization of the moment the tennis racket flips occurs. 

% \begin{figure}
%     \centering
%     \includegraphics[width=0.44\textwidth]{Figures/Visuals/output_grid.png}
%     \caption{3x3 grid indicating the moment of the tennis racket flip} 
%     \label{Fig:Tennis racket flip animation}
% \end{figure}

%\section{\label{sec:acknowledgements}acknowledgements}

%\ag{Figure, electric fields left, information fields right.}

%\ag{Andrew: Add details about this comparison with analytic expressions...}. 

%\ag{George/Seca: pyGDM results for spheres, hexagons}

\section*{Acknowledgements}
AAG is supported in part by NSF grants PHY-2110524 and PHY-2111544, the Heising-Simons Foundation, the John Templeton Foundation.  AAG and GW are supported by the W.M. Keck Foundation. AAG ZW and GW are supported by the DARPA levitas disruptioneering program. 
AAG, PJP, LF, ZF are supported by ONR Grant N00014-18-1-2370. 
GW is supported by a NASA NIAC phase 2 subaward from DRAPER labs. 
This work was supported in part by a grant from LeviNet. This work used the Quest high performance computing facility at Northwestern. We thank the Shariar group for a loan of a PPCL pure photonics laser and we thank Julian Gamboa for guidance on how to use it. We thank Professor Tracy Northrup of IQQI, Innsbruck, Austria, EU and Levinet (UK) for supporting Miriam M Florez's research time with the group. 
We thank the Barnard and Majumdar groups for lending us a Santec TSL-710 laser.

%\nocite{*}

\bibliography{apssamp}% Produces the bibliography via BibTeX.

@PREAMBLE{
 "\providecommand{\noopsort}[1]{}" 
 # "\providecommand{\singleletter}[1]{#1}%" 
}

@Article{Ju2023,
author={Ju, Peng
and Jin, Yuanbin
and Shen, Kunhong
and Duan, Yao
and Xu, Zhujing
and Gao, Xingyu
and Ni, Xingjie
and Li, Tongcang},
title={Near-Field GHz Rotation and Sensing with an Optically Levitated Nanodumbbell},
journal={Nano Letters},
year={2023},
month={Nov},
day={22},
publisher={American Chemical Society},
volume={23},
number={22},
pages={10157-10163},
issn={1530-6984},
doi={10.1021/acs.nanolett.3c02442},
url={https://doi.org/10.1021/acs.nanolett.3c02442}
}

@article{Rossi2025,
  title = {Quantum Delocalization of a Levitated Nanoparticle},
  author = {Rossi, M. and Militaru, A. and Carlon Zambon, N. and Riera-Campeny, A. and Romero-Isart, O. and Frimmer, M. and Novotny, L.},
  journal = {Phys. Rev. Lett.},
  volume = {135},
  issue = {8},
  pages = {083601},
  numpages = {7},
  year = {2025},
  month = {Aug},
  publisher = {American Physical Society},
  doi = {10.1103/2yzc-fsm3},
  url = {https://link.aps.org/doi/10.1103/2yzc-fsm3}
}

@article{ranfagni2022twodimensional,
  author = {Ranfagni, A. and B\o{}rkje, K. and Marino, F. and Marin, F.},
  doi = {10.1103/PhysRevResearch.4.033051},
  journal = {Phys. Rev. Res.},
  month = jul,
  number = 3,
  numpages = {10},
  pages = 033051,
  publisher = {American Physical Society},
  title = {Two-dimensional quantum motion of a levitated nanosphere},
  url = {https://link.aps.org/doi/10.1103/PhysRevResearch.4.033051},
  volume = 4,
  year = 2022
}

@article{kamba2022optical,
  author = {Kamba, Mitsuyoshi and Shimizu, Ryoga and Aikawa, Kiyotaka},
  doi = {10.1364/OE.462921},
  journal = {Opt. Express},
  month = jul,
  number = 15,
  pages = {26716--26727},
  publisher = {Optica Publishing Group},
  title = {Optical cold damping of neutral nanoparticles near the ground state in an optical lattice},
  url = {https://opg.optica.org/oe/abstract.cfm?URI=oe-30-15-26716},
  volume = 30,
  year = 2022
}

@article{tebbenjohanns2019damping,
  author = {Tebbenjohanns, Felix and Frimmer, Martin and Militaru, Andrei and Jain, Vijay and Novotny, Lukas},
  description = {Phys. Rev. Lett. 122, 223601 (2019) - Cold Damping of an Optically Levitated Nanoparticle to Microkelvin Temperatures},
  doi = {10.1103/PhysRevLett.122.223601},
  journal = {Phys. Rev. Lett.},
  month = jun,
  number = 22,
  numpages = {6},
  pages = 223601,
  publisher = {American Physical Society},
  title = {Cold Damping of an Optically Levitated Nanoparticle to Microkelvin Temperatures},
  url = {https://link.aps.org/doi/10.1103/PhysRevLett.122.223601},
  volume = 122,
  year = 2019
}

@article{delic2020cooling,
  author = {Deli{\'c}, Uro{\v s} and Reisenbauer, Manuel and Dare, Kahan and Grass, David and Vuleti{\'c}, Vladan and Kiesel, Nikolai and Aspelmeyer, Markus},
  doi = {10.1126/science.aba3993},
  issn = {0036-8075},
  journal = {Science},
  number = 6480,
  pages = {892--895},
  publisher = {American Association for the Advancement of Science},
  title = {Cooling of a levitated nanoparticle to the motional quantum ground state},
  url = {https://science.sciencemag.org/content/367/6480/892},
  volume = 367,
  year = 2020
}

@article{Colella:1975dq,
    author = "Colella, R. and Overhauser, A. W. and Werner, S. A.",
    title = "{Observation of gravitationally induced quantum interference}",
    doi = "10.1103/PhysRevLett.34.1472",
    journal = "Phys. Rev. Lett.",
    volume = "34",
    pages = "1472--1474",
    year = "1975"
}

@article{Werner:1979gi,
    author = "Werner, S. A. and Staudenmann, J. L. and Colella, R.",
    title = "{Effect of Earth's rotation on the quantum mechanical phase of the neutron}",
    doi = "10.1103/PhysRevLett.42.1103",
    journal = "Phys. Rev. Lett.",
    volume = "42",
    pages = "1103--1106",
    year = "1979"
}

@book{Rauch:2015jkh,
    author = "Rauch, Helmut and Werner, Samuel A.",
    title = "Neutron Interferometry: Lessons in Experimental Quantum Mechanics",
    doi = "10.1093/acprof:oso/9780198712510.001.0001",
    publisher = "Second Edition, Oxford University Press",
    year = "2015"
}

@article{Fixler:2007is,
    author = "Fixler, J. B. and Foster, G. T. and McGuirk, J. M. and Kasevich, M. A.",
    title = "{Atom interferometer measurement of the Newtonian constant of gravity}",
    doi = "10.1126/science.1135459",
    journal = "Science",
    volume = "315",
    pages = "74--77",
    year = "2007"
}

@article{Asenbaum:2016djh,
    author = "Asenbaum, Peter and Overstreet, Chris and Kovachy, Tim and Brown, Daniel D. and Hogan, Jason M. and Kasevich, Mark A.",
    title = "{Phase Shift in an Atom Interferometer due to Spacetime Curvature across its Wave Function}",
    eprint = "1610.03832",
    archivePrefix = "arXiv",
    primaryClass = "physics.atom-ph",
    doi = "10.1103/PhysRevLett.118.183602",
    journal = "Phys. Rev. Lett.",
    volume = "118",
    number = "18",
    pages = "183602",
    year = "2017"
}

@ARTICLE{Tonomura:1989,
       author = {{Tonomura}, A. and {Endo}, J. and {Matsuda}, T. and {Kawasaki}, T. and {Ezawa}, H.},
        title = "{Demonstration of single-electron buildup of an interference pattern}",
      journal = {American Journal of Physics},
         year = 1989,
        month = feb,
       volume = {57},
       number = {2},
        pages = {117-120},
          doi = {10.1119/1.16104},
       adsurl = {https://ui.adsabs.harvard.edu/abs/1989AmJPh..57..117T}
}

@article{Eibenberger:2013cqb,
    author = {Eibenberger, Sandra and Gerlich, Stefan and Arndt, Markus and Mayor, Marcel and T{\"u}xen, Jens},
    title = "{Matter-wave interference with particles selected from a molecular library with masses exceeding 10000 amu}",
    eprint = "1310.8343",
    archivePrefix = "arXiv",
    primaryClass = "quant-ph",
    doi = "10.1039/C3CP51500A",
    journal = "Phys. Chem. Chem. Phys.",
    volume = "15",
    pages = "14696--14700",
    year = "2013"
}

@article{Fein:2019dgf,
    author = "Fein, Yaakov Y. and Geyer, Philipp and Zwick, Patrick and Kia{\l}ka, Filip and Pedalino, Sebastian and Mayor, Marcel and Gerlich, Stefan and Arndt, Markus",
    title = "{Quantum superposition of molecules beyond 25 kDa}",
    doi = "10.1038/s41567-019-0663-9",
    journal = "Nature Phys.",
    volume = "15",
    number = "12",
    pages = "1242--1245",
    year = "2019"
}

@article{SGI_experiment,
  doi = {10.1126/sciadv.abg2879},
  author = {Yair Margalit and  Or Dobkowski and Zhifan Zhou and Omer Amit and Yonathan Japha and  Samuel Moukouri and Daniel Rohrlich and Anupam Mazumdar and Sougato Bose and Carsten Henkel and Ron Folman},
  title = {Realization of a complete Stern-Gerlach interferometer: Toward a test of quantum gravity},
  journal = {Science Advances},
  volume = {7},
  number = {22},
  pages = {eabg2879},
  year = {2021},
  URL = {https://www.science.org/doi/abs/10.1126/sciadv.abg2879}
}

@article{ORI11_GM,
   author = {Romero-Isart, Oriol},
   title = {Quantum superposition of massive objects and collapse models},
   journal = {Phys. Rev. A},
   volume = {84},   
   pages = {052121},
   DOI = {10.1103/PhysRevA.84.052121},
   url = {https://link.aps.org/doi/10.1103/PhysRevA.84.052121},
   year = {2011},
   type = {Journal Article}
}

@Article{Bateman2014,
author={Bateman, James
and Nimmrichter, Stefan
and Hornberger, Klaus
and Ulbricht, Hendrik},
title={Near-field interferometry of a free-falling nanoparticle from a point-like source},
journal={Nature Communications},
year={2014},
month={Sep},
day={02},
volume={5},
number={1},
pages={4788},
issn={2041-1723},
doi={10.1038/ncomms5788},
url={https://doi.org/10.1038/ncomms5788}
}

@article{Goldman2015,
  title = {Sensing short range forces with a nanosphere matter-wave interferometer},
  author = {Geraci, Andrew and Goldman, Hart},
  journal = {Phys. Rev. D},
  volume = {92},
  issue = {6},
  pages = {062002},
  numpages = {7},
  year = {2015},
  month = {Sep},
  publisher = {American Physical Society},
  doi = {10.1103/PhysRevD.92.062002},
  url = {https://link.aps.org/doi/10.1103/PhysRevD.92.062002}
}

@article{Schut:2024lgp,
    author = "Schut, Martine and Andriolo, Patrick and Toro{\v{s}}, Marko and Bose, Sougato and Mazumdar, Anupam",
    title = "{Expression for the decoherence rate due to air-molecule scattering in spatial qubits}",
    eprint = "2410.20910",
    archivePrefix = "arXiv",
    primaryClass = "quant-ph",
    doi = "10.1103/PhysRevA.111.042211",
    journal = "Phys. Rev. A",
    volume = "111",
    number = "4",
    pages = "042211",
    year = "2025"
}

@misc{pedalino2025probingquantummechanicsusing,
      title={Probing quantum mechanics using nanoparticle Schr\"odinger cats}, 
      author={Sebastian Pedalino and Bruno E. Ramírez-Galindo and Richard Ferstl and Klaus Hornberger and Markus Arndt and Stefan Gerlich},
      year={2025},
      eprint={2507.21211},
      archivePrefix={arXiv},
      primaryClass={quant-ph},
      url={https://arxiv.org/abs/2507.21211}, 
}

@article{Amit,
  title = {${T}^{3}$ Stern-Gerlach Matter-Wave Interferometer},
  author = {Amit, O. and Margalit, Y. and Dobkowski, O. and Zhou, Z. and Japha, Y. and Zimmermann, M. and Efremov, M. A. and Narducci, F. A. and Rasel, E. M. and Schleich, W. P. and Folman, R.},
  journal = {Phys. Rev. Lett.},
  volume = {123},
  issue = {8},
  pages = {083601},
  numpages = {6},
  year = {2019},
  month = {Aug},
  publisher = {American Physical Society},
  doi = {10.1103/PhysRevLett.123.083601},
  url = {https://link.aps.org/doi/10.1103/PhysRevLett.123.083601}
}

@article{Overstreet:2021hea,
    author = "Overstreet, Chris and Asenbaum, Peter and Curti, Joseph and Kim, Minjeong and Kasevich, Mark A.",
    title = "{Observation of a gravitational Aharonov-Bohm effect}",
    doi = "10.1126/science.abl7152",
    journal = "Science",
    volume = "375",
    number = "6577",
    pages = "abl7152",
    year = "2021"
}

@article{Kuhn:17,
author = {Stefan Kuhn and Alon Kosloff and Benjamin A. Stickler and Fernando Patolsky and Klaus Hornberger and Markus Arndt and James Millen},
journal = {Optica},
number = {3},
pages = {356--360},
publisher = {Optica Publishing Group},
title = {Full rotational control of levitated silicon nanorods},
volume = {4},
month = {Mar},
year = {2017},
url = {https://opg.optica.org/optica/abstract.cfm?URI=optica-4-3-356},
doi = {10.1364/OPTICA.4.000356},
}

@article{Li2018,
	author = {Ahn, Jonghoon and Xu, Zhujing and Bang, Jaehoon and Deng, Yu-Hao and Hoang, Thai M. and Han, Qinkai and Ma, Ren-Min and Li, Tongcang},
	doi = {10.1103/PhysRevLett.121.033603},
	issue = {3},
	journal = {Phys. Rev. Lett.},
	month = {Jul},
	numpages = {5},
	pages = {033603},
	publisher = {American Physical Society},
	title = {Optically Levitated Nanodumbbell Torsion Balance and GHz Nanomechanical Rotor},
	url = {https://link.aps.org/doi/10.1103/PhysRevLett.121.033603},
	volume = {121},
	year = {2018}}

@article{Xu2017Detecting,
  title = {Detecting Casimir torque with an optically levitated nanorod},
  author = {Xu, Zhujing and Li, Tongcang},
  journal = {Phys. Rev. A},
  volume = {96},
  issue = {3},
  pages = {033843},
  numpages = {12},
  year = {2017},
  month = {Sep},
  publisher = {American Physical Society},
  doi = {10.1103/PhysRevA.96.033843}
}

@article{rashid2018precession,
  title={Precession motion in levitated optomechanics},
  author={Rashid, Muddassar and Toro{\v{s}}, Marko and Setter, Ashley and Ulbricht, Hendrik},
  journal={Physical review letters},
  volume={121},
  number={25},
  pages={253601},
  year={2018},
  publisher={APS}
}

@article{winstone2022optical,
  title={Optical trapping of high-aspect-ratio NaYF hexagonal prisms for kHz-MHz gravitational wave detectors},
  author={Winstone, George and Wang, Zhiyuan and Klomp, Shelby and Felsted, Greg R and Laeuger, Andrew and Gupta, Chaman and Grass, Daniel and Aggarwal, Nancy and Sprague, Jacob and Pauzauskie, Peter J and others},
  journal={Physical review letters},
  volume={129},
  number={5},
  pages={053604},
  year={2022},
  publisher={APS}
}

@article{aggarwal2022searching,
  title={Searching for new physics with a levitated-sensor-based gravitational-wave detector},
  author={Aggarwal, Nancy and Winstone, George P and Teo, Mae and Baryakhtar, Masha and Larson, Shane L and Kalogera, Vicky and Geraci, Andrew A},
  journal={Physical Review Letters},
  volume={128},
  number={11},
  pages={111101},
  year={2022},
  publisher={APS}
}

@article{reimann2018ghz,
  title={GHz rotation of an optically trapped nanoparticle in vacuum},
  author={Reimann, Ren{\'e} and Doderer, Michael and Hebestreit, Erik and Diehl, Rozenn and Frimmer, Martin and Windey, Dominik and Tebbenjohanns, Felix and Novotny, Lukas},
  journal={Physical review letters},
  volume={121},
  number={3},
  pages={033602},
  year={2018},
  publisher={APS}
}

@article{petercrystalgrowth,
author = {Felsted, R. and Pant, Anupum and Bard, Alexander and Xia, Xiaojing and Luntz-Martin, Danika and Dadras, Siamak; zhang, Shuai and Vamivakas, Anthony and Pauzauskie, Peter },
title = {Chemically tunable aspect ratio control and laser refrigeration of hexagonal sodium yttrium fluoride upconverting materials},
journal = {Crystal Growth and Design},
year = {2022},
}

@article{ma2020quantum,
  title={Quantum persistent tennis racket dynamics of nanorotors},
  author={Ma, Yue and Khosla, Kiran E and Stickler, Benjamin A and Kim, MS},
  journal={Physical Review Letters},
  volume={125},
  number={5},
  pages={053604},
  year={2020},
  publisher={APS}
}

@article{westphal2021measurement,
  title={Measurement of gravitational coupling between millimetre-sized masses},
  author={Westphal, Tobias and Hepach, Hans and Pfaff, Jeremias and Aspelmeyer, Markus},
  journal={Nature},
  volume={591},
  number={7849},
  pages={225--228},
  year={2021},
  publisher={Nature Publishing Group UK London}
}

@article{gao2024feedback,
  title={Feedback cooling a levitated nanoparticle's libration to below 100 phonons},
  author={Gao, Jialiang and van der Laan, Fons and Zieli{\'n}ska, Joanna A and Militaru, Andrei and Novotny, Lukas and Frimmer, Martin},
  journal={Physical Review Research},
  volume={6},
  number={3},
  pages={033009},
  year={2024},
  publisher={APS}
}

@article{zeng2024optically,
  title={Optically levitated micro gyroscopes with an MHz rotational vaterite rotor},
  author={Zeng, Kai and Xu, Xiangming and Wu, Yulie and Wu, Xuezhong and Xiao, Dingbang},
  journal={Microsystems \& Nanoengineering},
  volume={10},
  number={1},
  pages={78},
  year={2024},
  publisher={Nature Publishing Group UK London}
}

@article{zielinska2024long,
  title={Long-Axis Spinning of an Optically Levitated Particle: A Levitated Spinning Top},
  author={Zieli{\'n}ska, JA and van der Laan, F and Norrman, A and Reimann, R and Frimmer, M and Novotny, L},
  journal={Physical Review Letters},
  volume={132},
  number={25},
  pages={253601},
  year={2024},
  publisher={APS}
}

@article{stickler2018probing,
  title={Probing macroscopic quantum superpositions with nanorotors},
  author={Stickler, Benjamin A and Papendell, Birthe and Kuhn, Stefan and Schrinski, Bj{\"o}rn and Millen, James and Arndt, Markus and Hornberger, Klaus},
  journal={New Journal of Physics},
  volume={20},
  number={12},
  pages={122001},
  year={2018},
  publisher={IOP Publishing}
}

@article{keith2010photophoretic,
  title={Photophoretic levitation of engineered aerosols for geoengineering},
  author={Keith, David W},
  journal={Proceedings of the National Academy of Sciences},
  volume={107},
  number={38},
  pages={16428--16431},
  year={2010},
  publisher={National Acad Sciences}
}

@article{rademacher2025roto,
  title={Roto-translational optomechanics},
  author={Rademacher, M and Pontin, A and Gosling, JMH and Barker, PF and Toro{\v{s}}, M},
  journal={arXiv preprint arXiv:2507.20905},
  year={2025}
}

@article{neumeier2024fast,
  title={Fast quantum interference of a nanoparticle via optical potential control},
  author={Neumeier, Lukas and Ciampini, Mario A and Romero-Isart, Oriol and Aspelmeyer, Markus and Kiesel, Nikolai},
  journal={Proceedings of the National Academy of Sciences},
  volume={121},
  number={4},
  pages={e2306953121},
  year={2024},
  publisher={National Academy of Sciences}
}

@book{nimmrichter2014macroscopic,
  title={Macroscopic matter wave interferometry},
  author={Nimmrichter, Stefan},
  year={2014},
  publisher={Springer}
}

@article{laplane2024inert,
  title={Inert shell coating for enhanced laser refrigeration of nanoparticles: application in levitated optomechanics},
  author={Laplane, Cyril and Ren, Peng and Roberts, Reece P and Lu, Yiqing and Volz, Thomas},
  journal={ACS Photonics},
  volume={11},
  number={3},
  pages={963--968},
  year={2024},
  publisher={ACS Publications}
}

@article{kuhn2017optically,
  title={Optically driven ultra-stable nanomechanical rotor},
  author={Kuhn, Stefan and Stickler, Benjamin A and Kosloff, Alon and Patolsky, Fernando and Hornberger, Klaus and Arndt, Markus and Millen, James},
  journal={Nature communications},
  volume={8},
  number={1},
  pages={1670},
  year={2017},
  publisher={Nature Publishing Group UK London}
}

@article{ahn2020ultrasensitive,
  title={Ultrasensitive torque detection with an optically levitated nanorotor},
  author={Ahn, Jonghoon and Xu, Zhujing and Bang, Jaehoon and Ju, Peng and Gao, Xingyu and Li, Tongcang},
  journal={Nature nanotechnology},
  volume={15},
  number={2},
  pages={89--93},
  year={2020},
  publisher={Nature Publishing Group UK London}
}

@article{laing2024optimal,
  title={Optimal displacement detection of arbitrarily-shaped levitated dielectric objects using optical radiation},
  author={Laing, Shaun and Klomp, Shelby and Winstone, George and Grinin, Alexey and Dana, Andrew and Wang, Zhiyuan and Widyatmodjo, Kevin Seca and Bateman, James and Geraci, Andrew A},
  journal={arXiv preprint arXiv:2409.00782},
  year={2024}
}

@article{rahman2017laser,
  title = {A laser cooled nanocryostat: Refrigeration, alignment and rotation of levitated Yb$^{3+}$:YLF nanocrystals},
  author = {Rahman, ATM and Barker, PF},
  journal = {arXiv preprint arXiv:1703.07155},
  year = {2017}
}

@article{van2017tennis,
  title={The tennis racket effect in a three-dimensional rigid body},
  author={Van Damme, L{\'e}o and Marde{\v{s}}i{\'c}, Pavao and Sugny, Dominique},
  journal={Physica D: Nonlinear Phenomena},
  volume={338},
  pages={17--25},
  year={2017},
  publisher={Elsevier}
}

@article{fleischer2022cryogenic,
  title={A cryogenic torsion balance using a liquid-cryogen free, ultra-low vibration cryostat},
  author={Fleischer, Svenja M and Ross, Michael P and Venkateswara, Krishna and Hagedorn, Charles A and Shaw, Erik A and Swanson, Erik and Heckel, BR and Gundlach, JH},
  journal={Review of Scientific Instruments},
  volume={93},
  number={6},
  year={2022},
  publisher={AIP Publishing}
}

@article{bang2020five,
  title={Five-dimensional cooling and nonlinear dynamics of an optically levitated nanodumbbell},
  author={Bang, Jaehoon and Seberson, Troy and Ju, Peng and Ahn, Jonghoon and Xu, Zhujing and Gao, Xingyu and Robicheaux, Francis and Li, Tongcang},
  journal={Physical Review Research},
  volume={2},
  number={4},
  pages={043054},
  year={2020},
  publisher={APS}
}

@article{kamba2023nanoscale,
  title={Nanoscale feedback control of six degrees of freedom of a near-sphere},
  author={Kamba, Mitsuyoshi and Shimizu, Ryoga and Aikawa, Kiyotaka},
  journal={Nature Communications},
  volume={14},
  number={1},
  pages={7943},
  year={2023},
  publisher={Nature Publishing Group UK London}
}

@article{pontin2023simultaneous,
  title={Simultaneous cavity cooling of all six degrees of freedom of a levitated nanoparticle},
  author={Pontin, Antonio and Fu, Hayden and Toro{\v{s}}, Marko and Monteiro, Tania S and Barker, Peter F},
  journal={Nature Physics},
  volume={19},
  number={7},
  pages={1003--1008},
  year={2023},
  publisher={Nature Publishing Group UK London}
}

@article{Xia2021a,
  title = {The Impact of {{2H}} {$\rightarrow$} {{4I}} Emission from {{Er3}}+ Ions on Ratiometric Optical Temperature Sensing with {{Yb3}}+/{{Er3}}+ Co-Doped Upconversion Materials},
  author = {Xia, Xiaojing and Volpi, Azzurra and Roh, Joo Yeon D. and De Siena, Michael C. and Gamelin, Daniel R. and Hehlen, Markus P. and Pauzauskie, Peter J.},
  year = 2021,
  month = aug,
  journal = {Journal of Luminescence},
  volume = {236},
  pages = {118006},
  issn = {00222313},
  doi = {10.1016/j.jlumin.2021.118006},
  urldate = {2025-11-03},
  langid = {english}
}

@article{Ivaturi2013,
  title = {Optimizing Infrared to near Infrared Upconversion Quantum Yield of {$\beta$}-{{NaYF4}}:{{Er3}}+ in Fluoropolymer Matrix for Photovoltaic Devices},
  shorttitle = {Optimizing Infrared to near Infrared Upconversion Quantum Yield of {$\beta$}-{{NaYF4}}},
  author = {Ivaturi, Aruna and MacDougall, Sean K. W. and {Mart{\'i}n-Rodr{\'i}guez}, Rosa and Quintanilla, Marta and {Marques-Hueso}, Jose and Kr{\"a}mer, Karl W. and Meijerink, Andries and Richards, Bryce S.},
  year = 2013,
  month = jul,
  journal = {Journal of Applied Physics},
  volume = {114},
  number = {1},
  pages = {013505},
  issn = {0021-8979, 1089-7550},
  doi = {10.1063/1.4812578},
  urldate = {2025-10-24},
  langid = {english}
}

@article{Renero-Lecuna2011,
  title = {Origin of the {{High Upconversion Green Luminescence Efficiency}} in {$\beta$}-{{NaYF}}{\textsubscript{4}} :2\%{{Er}}{\textsuperscript{3+}} ,20\%{{Yb}}{\textsuperscript{3+}}},
  shorttitle = {Origin of the {{High Upconversion Green Luminescence Efficiency}} in {$\beta$}-{{NaYF}}{\textsubscript{4}}},
  author = {{Renero-Lecuna}, C. and {Mart{\'i}n-Rodr{\'i}guez}, R. and Valiente, R. and Gonz{\'a}lez, J. and Rodr{\'i}guez, F. and Kr{\"a}mer, K. W. and G{\"u}del, H. U.},
  year = 2011,
  month = aug,
  journal = {Chemistry of Materials},
  volume = {23},
  number = {15},
  pages = {3442--3448},
  issn = {0897-4756, 1520-5002},
  doi = {10.1021/cm2004227},
  urldate = {2025-11-03},
  langid = {english}
}

@Article{Dania2025,
author={Dania, Lorenzo
and Kremer, Oscar Schmitt
and Piotrowski, Johannes
and Candoli, Davide
and Vijayan, Jayadev
and Romero-Isart, Oriol
and Gonzalez-Ballestero, Carlos
and Novotny, Lukas
and Frimmer, Martin},
title={High-purity quantum optomechanics at room temperature},
journal={Nature Physics},
year={2025},
month={Oct},
day={01},
volume={21},
number={10},
pages={1603-1608},
issn={1745-2481},
doi={10.1038/s41567-025-02976-9},
url={https://doi.org/10.1038/s41567-025-02976-9}
}

@article{Harris2020Array,
  title        = {Array programming with NumPy},
  author       = {Harris, Charles R. and Millman, K. Jarrod and van der Walt, St{\'e}fan J. and others},
  journal      = {Nature},
  volume       = {585},
  number       = {7825},
  pages        = {357--362},
  year         = {2020},
  doi          = {10.1038/s41586-020-2649-2},
  publisher    = {Nature Publishing Group}
}

@article{Virtanen2020SciPy,
  title        = {{SciPy} 1.0: Fundamental Algorithms for Scientific Computing in Python},
  author       = {Virtanen, Pauli and Gommers, Ralf and Oliphant, Travis E. and Haberland, Matt and Reddy, Tyler and Cournapeau, David and Burovski, Evgeni and Peterson, Pearu and Weckesser, Warren and Bright, Jonathan and van der Walt, St{\'e}fan J. and Brett, Matthew and Wilson, Joshua and Millman, K. Jarrod and Mayorov, Nikolay and Nelson, Andrew R. J. and Jones, Eric and Kern, Robert and Larson, Eric and Carey, CJ and Polat, {\.I}lhan and Feng, Yu and Moore, Eric W. and VanderPlas, Jake and Laxalde, Denis and Perktold, Josef and Cimrman, Robert and Henriksen, Ian and Quintero, E. A. and Harris, Charles R. and Archibald, Anne M. and Ribeiro, Ant{\^o}nio H. and Pedregosa, Fabian and van Mulbregt, Paul and {SciPy 1.0 Contributors}},
  journal      = {Nature Methods},
  volume       = {17},
  number       = {3},
  pages        = {261--272},
  year         = {2020},
  doi          = {10.1038/s41592-019-0686-2}
}

@article{Hunter2007Matplotlib,
  author       = {Hunter, J. D.},
  title        = {Matplotlib: A 2D Graphics Environment},
  journal      = {Computing in Science \& Engineering},
  volume       = {9},
  number       = {3},
  pages        = {90--95},
  year         = {2007},
  doi          = {10.1109/MCSE.2007.55}
}

@misc{NASA_Dzhanibekov_effect_2015,
  title        = {Dzhanibekov effect (video)},
  author       = {{NASA}},
  year         = {2015},
  howpublished = {\url{https://en.wikipedia.org/wiki/File:Dzhanibekov_effect.ogv}},
  note         = {Public domain (US government work)}
}

@article{hu2023structured,
  title={Structured transverse orbital angular momentum probed by a levitated optomechanical sensor},
  author={Hu, Yanhui and Kingsley-Smith, Jack J and Nikkhou, Maryam and Sabin, James A and Rodr{\'\i}guez-Fortu{\~n}o, Francisco J and Xu, Xiaohao and Millen, James},
  journal={Nature Communications},
  volume={14},
  number={1},
  pages={2638},
  year={2023},
  publisher={Nature Publishing Group UK London}
}

@article{arita2023cooling,
  title={Cooling the optical-spin driven limit cycle oscillations of a levitated gyroscope},
  author={Arita, Yoshihiko and Simpson, Stephen H and Bruce, Graham D and Wright, Ewan M and Zem{\'a}nek, Pavel and Dholakia, Kishan},
  journal={Communications Physics},
  volume={6},
  number={1},
  pages={238},
  year={2023},
  publisher={Nature Publishing Group UK London}
}

@article{millen2014nanoscale,
  title={Nanoscale temperature measurements using non-equilibrium Brownian dynamics of a levitated nanosphere},
  author={Millen, J and Deesuwan, T and Barker, P and Anders, Janet},
  journal={Nature nanotechnology},
  volume={9},
  number={6},
  pages={425--429},
  year={2014},
  publisher={Nature Publishing Group UK London}
}

@article{ranjit2015attonewton,
  title={Attonewton force detection using microspheres in a dual-beam optical trap in high vacuum},
  author={Ranjit, Gambhir and Atherton, David P and Stutz, Jordan H and Cunningham, Mark and Geraci, Andrew A},
  journal={Physical Review A},
  volume={91},
  number={5},
  pages={051805},
  year={2015},
  publisher={APS}
}

@article{LISAdrag,
title = {Gas damping force noise on a macroscopic test body in an infinite gas reservoir},
journal = {Physics Letters A},
volume = {374},
number = {34},
pages = {3365-3369},
year = {2010},
issn = {0375-9601},
doi = {https://doi.org/10.1016/j.physleta.2010.06.041},
url = {https://www.sciencedirect.com/science/article/pii/S0375960110007279},
author = {A. Cavalleri and G. Ciani and R. Dolesi and M. Hueller and D. Nicolodi and D. Tombolato and S. Vitale and P.J. Wass and W.J. Weber}
}

@article{ortiz2021laser,
  title={Laser Refrigeration by an Ytterbium-Doped NaYF4 Microspinner},
  author={Ortiz-Rivero, Elisa and Prorok, Katarzyna and Mart{\'\i}n, Inocencio Rafael and Lisiecki, Rados{\l}aw and Haro-Gonz{\'a}lez, Patricia and Bednarkiewicz, Artur and Jaque, Daniel},
  journal={Small},
  volume={17},
  number={46},
  pages={2103122},
  year={2021},
  publisher={Wiley Online Library}
}

\end{document}